\documentclass[a4paper]{article}
\usepackage{amssymb} 
\usepackage{amsmath}  
\usepackage{mathrsfs}
\usepackage{latexsym} 

\def\Z{\mathbb{Z}}
\def\sph {{}_{\sigma} Y _{j\mu}}
\def\nc {non commutative}
\def\gr {general relativity}

\def\spt {spacetime}

\def\Herm {\mbox{Herm}}
\def\Mat {\mbox{Mat}}

\def\End {\mbox{End}}
\def\calH {{\cal H}}

\def\calL {{\cal L}}
\def\calN {{\cal N}}
\def\calO {{\cal O}}

\def\ie {{i.e.}}

\def\setR{\mathbb{R}}

\def\setN{\mathbb{N}}
\def\setC{\mathbb{C}}

\def\setZ {\mathbb{Z}}

\def\setK{\mathbb{K}}
\def \guill {\textquotedblleft}
\newcommand{\ket}[1]{\mid #1 \,\rangle}
\newcommand{\bra}[1]{\langle \,#1 \mid}
\newcommand{\braket}[2]{\langle \, #1 \mid #2 \,\rangle}
\newcommand{\sss}[1]{\scriptscriptstyle #1}

\newcommand{\Id}[1]{\mathrm{Id}_{#1}}

\title{Fuzzy spheres from inequivalent  coherent states  quantizations}
\author{Jean Pierre Gazeau, Eric Huguet, Marc Lachi\`eze-Rey and Jacques Renaud\\
\emph{Laboratoire
Astroparticules et Cosmologie$^{\ddag}$}
\\  \emph{Boite 7020, Universit\'e Paris 7-Denis Diderot}�\ \emph{F-75251 Paris Cedex 05, France}\\
\small{gazeau@ccr.jussieu.fr, 
huguet@ccr.jussieu.fr, marclr@cea.fr, renaud@ccr.jussieu.fr}\\
{\footnotesize{}$^{\ddag}$UMR 7164 (CNRS,Universit\'e Paris 7, CEA, Observatoire de Paris)}}
\date{July 4, 2006}
\begin{document}
\maketitle

\abstract{We present a new procedure which allows a coherent state (CS) quantization of any set with a measure. It  is manifest through the  replacement of  classical observables by  CS quantum observables, which acts on a Hilbert space of prescribed dimension $N$. The algebra of  CS quantum observables has the  finite dimension $N^2$.

The application  to the 2-sphere provides  a   family of inequivalent  CS  quantizations,   based on the  spin spherical harmonics (the CS quantization from usual spherical harmonics appears to give a trivial issue for the cartesian coordinates). We compare these   CS  quantizations  to the usual  (Madore) construction  of the fuzzy sphere. The difference allows us to consider our  procedures as the constructions of  new type of fuzzy spheres.  The very general character  of our method  suggests applications to construct fuzzy versions of a variety of sets. }

%%%%%%%%%%%%%%%%%%%%%INTRODUCTION%%%%%%%%%%%%%%%%%%

\section{Some ideas on quantization}

A classical description of a set of data, say $X$, is usually carried out  by
considering sets of  real or complex functions on $X$. Depending on the
context (data handling, signal analysis, mechanics\ldots)
the set $X$ will be equipped with a definite structure (topological space,
measure space, symplectic manifold\ldots) and the set of functions
on $X$ which will be considered as \emph{classical observables} must be restricted with regard to the structure on $X$; for instance,    signals should be
square integrable with respect to the  measure assigned to set $X$.

How  to provide instead a ``quantum description" of the 
same set $X$?  As a first characteristic, the latter replaces -  this 
is a  definition - the classical observables by   \emph{quantum 
observables}, which do not  commute in general. As usual, these quantum 
observables will be realized as operators acting on some Hilbert space 
$\calH$, 
whose projective version will be considered as the set of quantum 
states. This Hilbert space  will be constructed   as a subset in the set of   
functions on $X$.

The advantage  of the  coherent states (CS)  quantization 
procedure, in a standard sense \cite{klau1,ber,klau2} as in  recent generalizations \cite{gghlrl} and applications \cite{GPW} is that it requires  a minimal significant structure   on 
$X$, namely the only existence of a measure
$\mu(dx)$, together with a $\sigma$-algebra of measurable subsets. As
a measure space, $X$ will be given the name of an {\it observation}
set in the present context, and the existence of a measure provides
us with a statistical reading of the set of measurable real or
complex valued functions on $X$: computing for instance
average values on subsets with bounded measure. The  quantum states will 
correspond  to  measurable and square integrable functions on the set $X$,
but not all square integrable functions are eligible as quantum states.
The construction of  $\calH$ is equivalent to   the choice of a class 
of eligible quantum states, together with a technical condition of continuity. 
This provides
a correspondence between classical and quantum observables by
defining a generalization of the so-called coherent states.

Although    the procedure appears mathematically  as a quantization, 
it may also  be considered as a change of  point of view for  
looking at the system, not necessarily  a path to quantum physics. In this 
sense,   it could be called a discretization or a regularization \cite{Taylor}.
It shows a  certain resemblance with standard
procedures pertaining  to signal processing, for instance those involving
wavelets, which are   coherent states for the affine group transforming the half-plane time-scale into itself \cite{daube,csbook}. In many
respects, the choice of a quantization appears here as the choice of
a resolution to look at the system. 

As  is well known, some aspects of (ordinary) quantum mechanics may 
be seen as a \nc ~version of the geometry of the phase space, where position 
and 
momentum operators do not commute. It appears as  a  general fact that the 
quantization of  
a ``set of data'' makes   a fuzzy (\nc) geometry to  emerge \cite{Madore}.  We will show 
explicitly
how the CS  quantization of the ordinary sphere leads to its fuzzy geometry.
 
 In  Section 2  we present a  construction of coherent states which is very general and encompasses most of the
known constructions, and we derive from the existence of a CS family what we call CS quantization. The latter
extends to various situations the well-known Klauder-Berezin quantization. The formalism is illustrated with 
the standard Glauber-Klauder-Sudarshan coherent states and the related canonical quantization of the classical
phase space of the motion on the real line. 
 
 In Section 3, we apply the formalism to the sphere $S^2$ by using orthonormal families of spin spherical harmonics $\left( {}_{\sigma}Y_{jm}\right)_{-j \leqslant m \leqslant j}$ \cite{Newman,gmnrs,Campbell}. For a given $\sigma$ such that $2\sigma \in \setZ$ and $j$ such that $2 \vert\sigma\vert \leqslant 2j \in \setN$ there corresponds a continuous family of coherent states and the subsequent $2j+1$-dimensional quantization of the 2-sphere.  For a given $j$, we thus get $2j+1$ \underline{inequivalent} quantizations, corresponding to the   possible values of $\sigma$.   Note that the classical Gilmore-Perelomov-Radcliffe case \cite{Gilmore, Radcliffe,per} correspond to the particular value $\sigma = j$. On the other hand, the case $\sigma = 0$ is proved to be singular in the sense that it leads to a null quantization of the cartesian coordinates of the 2-sphere.
 
 The section 4 establishes the link between the CS quantization approach to the 2-sphere and the Madore construction \cite{Madore,gghlrl1} of the fuzzy sphere. We examine  there the question of equivalence between the two procedures. Note that 
a construction of  the fuzzy sphere based on Perelomov coherent states  has already been carried out by Grosse and Pre\u{s}najder \cite{gropre}. They proceed to  a covariant symbol calculus \emph{\`a la} Berezin with its corresponding $\star$-product. However, their approach  is different of ours.
 
 The appendices give an exhaustive set of formulas, particularly concerning the spin spherical harmonics,  needed for a complete description of our CS approach to the 2-sphere.

%%%%%%%%%%%%%%%%%%%%%SECTION2%%%%%%%%%%%%%%%%%%%%%

\section{Coherent states}
\subsection{The construction}\label{construction}
The	(classical) system to be quantized is considered as a set
of data, $X = \{x \in X\}$, assumed to be equipped with a measure $\mu$ 
defined on a
$\sigma$-field $\cal B$. We consider the Hilbert spaces 
${\rm L}_{{\setK}}^{2}(X, \mu)$ ($\setK=\setR\mbox{ or }\setC$) of real or 
complex functions, 
with the usual Hermitian inner product $\langle f \mid  g\rangle$. The 
quantization 
is defined by the choice of a closed subspace $\calH$ of ${\rm 
L}_{{\setK}}^{2}(X,d\mu)$. 
The only requirements on $\calH$, in addition  to be an Hilbert space, amount to the
following technical conditions:  

\begin{itemize}
\item{\sl
For all $\psi \in \mathcal{H}$ and all $x$, $\psi(x)$ is well defined (this is of course the case whenever $X$ is a topological space and the elements of $\cal H$ are continuous functions)
\item 
the linear map (``evaluation map'')
%\begin{equation}\label{qmap}
%\delta_x\ \begin{array}{rcl}
%{\cal H}&\to&\setK\\
%\psi&\mapsto&\psi(x)\end{array}
%\end{equation}
\begin{align}
\delta_x : &~{\cal H} \to \setK \label{qmap}\\
&~\psi \mapsto \psi(x)\nonumber
\end{align}
is continuous with respect to the topology of $\cal H$, for almost all $x$.}
\end{itemize}

The last condition is   realized as soon as the space
$\cal H$ is finite dimensional since  all the linear forms are continuous in 
this case. We see below that some other examples can be found.

As a consequence, using the Riesz theorem, there exists, for almost all $x$,  an 
unique
element $p_x \in\calH$  (a function) such that
\begin{equation}\label{CSmap1}
\langle p_x \mid \psi\rangle=\psi(x).
\end{equation}
We define the {\sl  coherent states} as the normalized vectors corresponding 
to $p_x$ written in
Dirac notation:
\begin{equation}\label{CS1}
 \ket{x} \equiv  \frac{\ket{p_x}}{[{\cal N}(x)]^{\frac{1}{2}}}\mbox{ where }
\calN (x) \equiv \langle p_x\mid p_x\rangle.
\end{equation} 
One can see at once that, for any $\psi\in\calH$: 
\begin{equation}\label{CSmap2}
\psi(x)=\left[\calN(x)\right]^{\frac{1}{2}}\langle
x\mid\psi\rangle.
\end{equation}
As a consequence, one obtains the following
resolution of the identity of $\calH$ which is at the basis of the whole 
construction:
\begin{equation}\label{resid}
\Id{\calH}=\int\ket{x}\bra{x}\calN(x)~ \mu(dx).
\end{equation}
Note that
\begin{equation}
 \phi(x) = \int_X \sqrt{\mathcal
N(x)~\mathcal N(x')}~ \langle x| x' \rangle ~\phi(x')\, \mu(dx'),~\forall \phi 
\in   {\mathcal H}. 
\end{equation}
Hence, ${\mathcal H}$ is     a reproducing Hilbert space with kernel 
\begin{equation} 
 K(x,x') =
\sqrt{\mathcal N(x)~\mathcal N(x')} ~\langle x\, | x' \rangle,
\end{equation} 
and the latter assumes finite diagonal values ({\it
a.e.}), ${K}(x,x) = \mathcal N(x)$, by construction.
Note that this construction yields an embedding of $X$ into $\cal
H$ and one could interpret $\ket{x}$ as a state localized at $x$ once a notion of localization has been properly defined on $X$.

In view of (\ref{resid}) the set $\{\ket{x}\}$ is called a {\sl  frame} 
for $\cal H$. This frame is said to be overcomplete when the 
the vectors $\{\ket{x}\}$ are not linearly independent \cite{aag,AEG}.

We define a \emph{classical observable} over $X$ in a loose way as a function $f:X \mapsto \setK$ ($\setR$
or $\setC$). As a matter of fact  
 we will not retain \emph{a priori} the   
usual requirements  on $f$ like to be real valued and   smooth with respect to  some 
topology   defined on $X$.

To any such function $f$, we associate the quantum observable over
${\calH}$ through the map:  
\begin{equation}\label{quant} 
f \mapsto A_{f} \equiv \int_X~
{\cal N}(x)~\mu(dx)~f(x)~ \, \ket{x} \bra{x} . 
\end{equation} 
The  operator  corresponding to a real function is Hermitian by construction. 
Hereafter,  we will also  use the notation $\tilde{f}$ for $A_f$.

The existence of the continuous frame $\{\ket{x}\}$ enables us to 
carry out  a symbolic calculus {\it \`a la} Berezin-Lieb
\cite{ber,csfks}. To each linear, self-adjoint operator (observable) ${\cal
O}$ acting on
$\calH$, one associates the {\sl lower (or covariant) symbol}
\begin{equation} \label{lsymb}
\check{\cal O}(x)\,\equiv \,\bra{x} {\cal O} \ket{x}, 
\end{equation}
and the {\sl upper (or contravariant) symbol} (not necessarily
unique) $\widehat{\cal O}$ such that
\begin{equation}\label{usymb} {\cal O} \,=\,\int_X ~{\cal N}(x)~\mu(dx) 
~\widehat{\cal
O}(x) \ket{x} \bra{x}.
\end{equation}
Note that $f$ is an upper symbol of $A_{f}$.

The technical conditions  and the definition of coherent states can be easily 
expressed when we have a  Hilbertian basis of ${\cal H}$.
Let $(\phi_n)_{ n\in I}$ such a basis, the technical condition
is equivalent to 
\begin{equation}
\sum_n|\phi_n(x)|^2<\infty\ \mbox{a.e.}
\end{equation}
The coherent state is then defined by
$$| x\rangle=\frac{1}{\left({\cal N}(x)\right)^{\frac{1}
{2}}}\sum_n\phi_n^*(x)~\phi_n\mbox{ with }
{\cal N}(x)= \sum_n|\phi_n(x)|^2.$$
To a certain extent, the
quantization scheme exposed here consists in adopting a certain point 
of view in dealing with $X$,  determined by the choice of the space $\cal H$. 
This choice specifies the admissible quantum states and the correspondence 
``classical observables versus quantum observables'' follows.

\subsection{The standard coherent states} 
Let us illustrate the above construction for the   dynamics of a  particle 
moving on the real line. This leads to  the well-known 
Klauder-Glauber-Sudarshan  coherent states \cite{klauskag} and the subsequent 
so-called {\sl canonical quantization} (with a slight difference of notation). 
The construction can be easily extended to the dynamics of the  particle in a 
flat higher dimensional  \spt. The observation set $X$ is the classical phase 
space $\setR^2 \simeq \setC = \{ z =\frac{1}{\sqrt2}(q+ip) \}$ (in complex 
notations) of a particle with
one degree of freedom. The symplectic form  identifies with $\frac{i}{2} 
~dz\wedge d\bar z \equiv d^2z$, the Lebesgue measure of the plane.  Here we 
adopt the   Gaussian   measure on $X$, $\mu(dz) = \frac{1}{\pi}\, e^{-\vert z 
\vert^2}\, d^2z $.

The quantization of $X$   is hence achieved by  a \emph{choice of 
polarization} (in the language of   geometric quantization): the  selection,  
in ${\rm L}^2(X,d\mu)$,  of  the Hilbert subspace  ${\mathcal H}$ defined as 
the so-called Fock-Bargmann space  of    all antiholomorphic entire functions 
that are square integrable with respect to the Gaussian measure.  

The Hilbertian basis is given by the  functions $\phi_n (z) \equiv 
\frac{\bar{z}^n}{\sqrt{n!}}$,  the normalized powers of the conjugate of the complex variable 
$z$. Thus, since $
\sum_n \frac{\vert z \vert^2}{n!} = e^{\vert z \vert^2}$, the coherent states 
read
\begin{equation}
| z\rangle = e^{-\frac{\vert z \vert^2}{2}} \sum_n
\frac{  z^n}{\sqrt{n!}}| n\rangle, \label{scs}
\end{equation} 
where $| n\rangle$ stands for $\varphi _n$, and one easily checks the 
normalization and unity resolution:
\begin{equation}
\langle z\, | z \rangle = 1,\ \
\frac{1}{\pi}\int_{\setC} | z\rangle \langle z| \, d^2 z= \Id{{\mathcal H}}. 
\label{pscs}
\end{equation}
Note that the reproducing kernel is simply given by $K(z,z')=e^{z\bar{z}'}$.

Quantum operators acting on ${\mathcal H}$ are yielded by using (\ref{quant}). 
We thus have for the most basic one,
\begin{equation}
a \equiv A_z = \frac{1}{\pi}\int_{\setC} z\, | z\rangle \langle z| \,
d^2z  = \sum_n \sqrt{n+1}~ | n\rangle \langle n+1|,
\label{low} 
\end{equation}       
which appears as the lowering operator, $a | n\rangle =
\sqrt{n}~ | n - 1\rangle$. Its adjoint
$a^{\dagger}$ is obtained by replacing $z$ by $\bar{z}$ in
(\ref{low}), and we get the factorization $N = a^{\dagger}a$ for the number 
operator, together
with the commutation rule $\lbrack a, a^{\dagger} \rbrack = \Id{\calH}$. Also 
note that $ a^{\dagger}$ and $a$ realize on ${\mathcal H}$ as
multiplication operator and derivation operator respectively,
$a^{\dagger}f(z) = zf(z), \ af
= df/dz$. From $q = \frac{1}{\sqrt{2}}(z +
\bar{z})$ et $p = \frac{1}{\sqrt{2}i}(z - \bar{z})$, one easily infers by
linearity that $q$ and $p$
are upper symbols for $\frac{1}{\sqrt{2}}(a + a^{\dagger}) \equiv Q$ and
$\frac{1}{\sqrt{2}i}(a - a^{\dagger})
\equiv P$ respectively. In consequence, the (essentially) self-adjoint operators
$Q$ and $P$ obey the canonical
commutation rule $\lbrack Q, P \rbrack = i\Id{\calH}$, and for
this reason fully deserve the
name of position and momentum operators of the usual (Galilean)
quantum mechanics, together with
all localization properties specific to the latter. 

%%%%%%%%%%%%%%%%%%%%%SECTION3%%%%%%%%%%%%%%%%%%%%%

\section{Quantizations of the 2-sphere}

\subsection{The 2-sphere}

We now  apply our method to the quantization of the observation
set $X =S^{2}$, the unit 2-sphere. This is   not to be confused with the 
quantization of  the phase space for the motion  on the two-sphere (\ie  quantum 
mechanics   on the two-sphere, see for instance  \cite{Kowalski}, 
\cite{Kowalski2}, \cite{hall}).  A 
point of $X$ is denoted by its spherical coordinates, $x =(\theta,\phi)$. Through the usual embedding in 
$\setR^{3}$, we may see 
$x$ as a point $ (x^i) \in \setR^{3}$ obeying
$\sum _{i=1}^3(x^i)^2=1$. 
We adopt on $S^2$  the normalized measure $\mu (dx) = 
\sin \theta ~d\theta~d\phi/4\pi$, proportional to the SO(3)-invariant
measure, which is also the surface element.

We know that $\mu$ is a symplectic form, with the canonical coordinates 
$q=\phi,\, p=-\cos
\theta$. This allows to see $S^{2}$ itself as the phase space for the theory
of (classical) angular momentum. In this spirit, we will be able to
interpret our procedure as the construction of families of {\sl  spin coherent 
states} including the Gilmore-Perelomov-Radcliffe (hereafter, GPR) ones \cite{per}. Also, our construction will take advantage of the group
action of SO(3) on $S^2$  embedded in~$\setR^{3}$. This three-dimensional   
group   acts as isometries in 
~$\setR^{3}$, as rotations in $S^{2}$. However, we emphasize again that
our quantization  procedure is based on the only existence of a
measure, and may be used in the absence of metric or symplectic
structure.

\subsection{The CS quantization of the 2-sphere}\label{Hermitian processing}

\subsubsection{The Hilbert space and the coherent  states}

At the basis of the CS  quantization procedure  is the choice of a finite dimensional Hilbert space, which is a subspace of  $L^2(S^2)$, and which carries a UIR of the group SU(2).  
We  write its dimension $2j+1$, with $j$ integer or semi-integer.
Although it could have appeared natural to choose this space as $V^j$, the linear span of ordinary spherical harmonics $Y_{jm}$, this choice would  not allow to consider half-integer values of $j$. Moreover, it happens that the quantization so obtained gives trivial results for the cartesian coordinates. Namely, the quantum counterparts of the cartesian coordinates (or, equivalently, the  spherical harmonics $Y_{1m}$) are identically zero. 
 Thus we are led to define $\calH$ on a general setting as  the  linear span of
spin spherical harmonics (hereafter SSH's).

\subsubsection{The spin spherical harmonics}
  
  We define $\calH=\calH ^{\sigma j}$ as the vector space spanned by   the spin spherical harmonics 
$   _{\sigma} Y_{j\mu} \in L^2(S^2)$, where $-j\le \sigma, \mu \le j$, and $\sigma$ is fixed in this range. Note that $\sigma$ and $j$ are both  integer or semi-integer. The spin spherical harmonics (SSH's) were   first 
introduced in \cite{Newman} (see also \cite{Campbell} and \cite{gmnrs} for their main properties). In view of their importance in the context of the present work, they are comprehensively described in Appendix A.   
The   special case $\sigma =0$ corresponds to the ordinary spherical harmonics $$ _{0} Y _{jm}= Y _{jm}.$$

A CS quantization is defined after a choice of values for   $j$ and $\sigma$, that we consider as  fixed in the sequel.
With the usual inner product of $L^2(S^2)$, the SSH's provide an  ON basis $(\sph)_{\mu=-j...j}$ of $\calH ^{\sigma j}$ (hereafter the SSH basis). 

The Hilbert space $\calH ^{\sigma j}$ carries the $2j + 1$-dimensional UIR of SU(2) (see Appendix A).
The generators of SU(2) in this representation can be taken as those corresponding to the three rotations around the orthogonal axes of $x^1,x^2,x^3$.
They 
are called the  ``spin'' angular momentum operators (SAMOs, to be distinguished from the usual angular momentum operators $J_i$), and will be written as  $\Lambda^{\sigma j}_a$. Hereafter, the index  $a=1,2,3$ will refer to the three spatial directions. We have  
$\Lambda ^{0 j}_a=J_a$, the usual angular momentum operators.  As usual, we define   $\Lambda ^{\sigma j}_\epsilon = \Lambda ^{\sigma j}_1 +\epsilon ~i~ \Lambda ^{\sigma j}_2,~\epsilon=\pm 1$.
All these generators   obey the usual commutation relations of the group  SU(2). 
They act on the ON basis as
\begin{equation}
\label{Lambdai}
\Lambda ^{\sigma j}_3 ~\sph=\mu~\sph,~
\Lambda ^{\sigma j}_\epsilon ~\sph=a_\epsilon (j,\mu)~ _{\sigma} Y _{j\mu +\epsilon},\end{equation}
where  the  $a_\epsilon (j,\mu) $,  given in (\ref{angspher+},\ref{angspher-}),  are the same as for the usual angular momentum   operators $J_a$.

The SSH  basis allows to  identify
$\calH ^{\sigma j}$  with $\setC^{2j+1}$: 
\begin{equation}
 \sph\, \leadsto \,\ket{\mu} \, \hookrightarrow \,(0,\ldots,0,1,0,
\ldots,0)^t~
\mathrm{with}~ \mu = -j,-j+1, \ldots, j\;, 
\end{equation}
where the 1 is at position $\mu$ and the superscript $t$ denotes the transpose. By construction we have the  Hilbertian orthonormality  relations:
\begin{equation}
 \braket{\mu}{\nu}\equiv \int_X \mu(dx) ~_{\sigma} Y _{j\mu} ^*(x) ~_{\sigma} Y _{j\nu} (x) = \delta_{\mu \nu}.
\end{equation}

The CS construction presented in Sect. (\ref{construction}) leads to the following class of 
coherent  states
\begin{equation} \ket{x} \,=\ket{ \theta, \phi}= \frac{1}{\sqrt{\calN (x)}}  
  \sum _{\mu=-j}^j 
~ _{\sigma} Y _{j\mu}^* (x)~\ket{\mu} ;~\ket{x} \in \calH,
\end{equation}
with $$\calN (x)=\sum _{\mu=-j}^j  \mid  _{\sigma} Y _{j\mu} (x)\mid  ^2=\frac{2j+1}{4\pi}.$$

For $\sigma = \pm j$, they reduce to   the  \emph{spin coherent states} \cite{Gilmore,Radcliffe,per}.

\subsubsection{Operators}

 We call ${\cal O }^{\sigma j}\equiv \End({\cal H }^{\sigma j})$ the space of linear operators (endomorphisms) acting on ${\cal H }^{\sigma j}$. This is a complex vector space of dimension $(2j+1)^2$ and an algebra for the natural composition of endomorphisms.
The 
SSH basis  allows to write a  linear endomorphism of $\calH ^{\sigma j}$ (\ie, an element of $\calO ^{\sigma j}$)   in a matrix form.  This provides   the  algebra isomorphism 
$$\calO ^{\sigma j}  \leadsto \Mat_{2j+1},$$ the algebra of complex matrices  of order $2j+1$, equipped with the matrix  product.

The  projector 
$ \ket{x} \bra{x}$ is  a particular  linear endomorphism of  $\calH ^{\sigma j} $, \ie, an element of $\calO ^{\sigma j}$.  
Being    Hermitian  by construction, it 
may be seen as  an Hermitian 
matrix of order $2j+1$, \ie, an element of 
  $\Herm_{2j+1} \subset    \Mat_{2j+1}$. Note that $\Herm_{2j+1}$  and $ \Mat_{2j+1}$ have respective (complex) 
dimensions $(j+1)~(2j+1)$ and $(2j+1)^2$. 

We have  resolution of identity and normalization by construction: 
\begin{equation} ~\int _{S^{2} } \mu(dx)~ \calN (x)\,\ket{x} \bra{x}\,=\,\Id,
\hspace{2cm}
\braket{x}{x}\,=\,1.\nonumber
\end{equation}

\subsubsection{Observables}

According to the prescription (\ref{quant}),  the CS quantization associates to the 
\emph{classical} observable $f:~S^2 \mapsto \setC$  the \emph{quantum}  observable  
\begin{align}\label{ccvv}
\nonumber \tilde{f} &\equiv  A_f = \int \mu(dx)~f(x)~\mathcal{N}(x)~
\ket{x} \bra{x}\\
 &= \sum _{\mu,\nu=-j}^{j}~\int \mu(dx)~f(x) 
~[ \sph (x)]^{\ast}  ~  _{\sigma} Y _{j \nu} (x)~
\ket{\mu} \bra{\nu}.
\end{align} 
This operator is an element of    ${\cal O }^{\sigma j} \sim \End (\calH ^{\sigma j}) \leadsto \mbox{Mat}_{(2j+1)}$. Of course its existence is submitted to the convergence of (\ref{ccvv}) in the weak sense  as an  operator integral. 
The expression above gives directly its expression as a matrix in the SSH basis, with matrix elements
$\tilde{f}_{\mu \nu}$: 
\begin{equation} 
\tilde{f} = \sum _{\mu,\nu=-j}^{j} \tilde{f}_{\mu \nu} ~\ket{\mu} \bra{\nu}
~\mathrm{with}~\tilde{f}_{\mu \nu}= \int \mu(dx)~f(x) ~  _\sigma Y_{j\mu}^*(x) ~ _\sigma Y_{j\nu}(x). 
\end{equation}

When $f$ is 
real-valued, the corresponding matrix  belongs to $\mbox{Herm}_{(2j+1)}$.
Also, we have $\widetilde{f^*}  = (\tilde{f} )^\dag$ (matrix  transconjugate), where we have used the same notation for the operator and the associated matrix.

\subsubsection{The usual spherical harmonics as classical observables}

 An  usual spherical harmonics $Y  _{\ell  m}$ is a  particular classical observable and, as such, may be quantized.
The quantization procedure  associates to $Y  _{\ell  m}$  the operator  $\widetilde{Y  _{\ell  m}}$. The details of the computation are given in Appendix A and the result is given in Subsection \ref{spherquant}, Eq. (\ref{quantsphergen}). We hence obtain 
the   matrix elements of $\widetilde{Y  _{\ell  m}}$  in the SSH  basis:
\begin{equation}\label{MAT}
\left[\widetilde{Y  _{\ell  m}}\right]_{\mu\nu}= 
(-1)^{\sigma -\mu} ~(2j  + 1)~  \sqrt{\frac{(2\ell + 1) }{4 \pi}} ~ 
  \begin{pmatrix}
j  & j  & \ell \\
 -\mu  & \nu & m
\end{pmatrix}  \begin{pmatrix}
 j  & j & \ell  \\
-\sigma  & \sigma & 0
\end{pmatrix},
\end{equation}
in terms of  the $3j$-symbols. This generalizes   
the  formula (2.7) of \cite{Freidel}. This expression  is a real quantity.

Any function $f$ on the 2-sphere with reasonable properties (continuity,  integrability...) may be expanded  in spherical harmonics as 
\begin{equation}
f=\sum _{\ell=0}^{\infty} ~\sum _{m=-\ell}^\ell~ f_{\ell m} ~Y _{\ell 
m},
\end{equation} 
from which  results the corresponding  expansion of $\tilde{f}$. However, the   
$3j$-symbols  are non zero only when a triangular inequality is satisfied. This  implies   that  the expansion   is cut at a finite value, 
giving  
\begin{equation}
\tilde{f} = ~\sum _{\ell = 0}^{2j} \sum _{{\sss m}=-\ell}^{\ell} ~f_{\ell {\sss 
m}} ~\widetilde{ Y_{\ell{\sss m}}}.
\end{equation}
This relation means that the  $(2j+1)^2$ observables $(\widetilde{Y  _{\ell  
m}})_{\ell \leqslant 2j,
~-\ell \leqslant m \leqslant \ell}$
  provide   a  second (SH) basis of 
$ {\cal O}^{\sigma j}$. 

The $f_{\ell {\sss  m }}$ are the components of the matrix $ \tilde{f}  \in {\cal O}^{\sigma j}$ in this basis.

\subsection{The spin angular momentum operators}

\subsubsection{Action on functions}

The  Hilbert space   $\calH ^{\sigma j}$ carries a unitary irreducible representation of the group SU(2) with   generators   $\Lambda ^{\sigma j}_a$ (the SAMOs), which belong to ${\cal O}^{\sigma j}$.
Their action    is  given in 
(\ref{angspher}-\ref{angspher+}-\ref{angspher-}).
Explicit calculations  shown in the appendix (see \ref{identvectam}) give the crucial relations:
\begin{equation}\label{identvectam2}
 \widetilde{x^a}  = K~ \Lambda ^{\sigma j}_a ,~\mathrm{with}~K\equiv \frac{\sigma}{j(j+1)}. 
 \end{equation}
 
We see here the peculiarity of the ordinary spherical harmonics ($\sigma =0$) as an ON basis for the quantization procedure: they would lead to a trivial result for the quantized version of the cartesian coordinates! On the other hand, the quantization based on the GPR spin coherent states yields the maximal value: $K = 1/(j+1)$. 
  Hereafter we assume $\sigma \ne 0$.
  
\subsubsection{Action on operators}

The SU(2)  action on $\calH ^{\sigma j}$  induces the following  
canonical (infinitesimal) action   on  ${\cal O} ^{\sigma j} =\End(\calH ^{\sigma j}) $: 
\begin{equation}
\calL ^{\sigma j}_a :~\mapsto \calL ^{\sigma j} _a A \equiv [\Lambda ^{\sigma j}_a, A] ~~\mbox{(the~commutator)}
\end{equation} here expressed through the generators.

We  prove in Appendix A,  (\ref{infcomm}), that 
%\begin{equation*}
$\calL ^{\sigma j} _a~ \widetilde{ Y_{\ell m}}  =\widetilde{   J_a  Y_{\ell m}}$ ,
%\end{equation*}
from which it results:   
\begin{equation*}
\calL ^{\sigma j}_3  \widetilde{ Y_{\ell m}}   
=m~\widetilde{  Y_{\ell m}} \mbox{~and~} (\calL^{\sigma j}) ^2  \widetilde{ Y_{\ell m}}   
%=\widetilde{(J^{(\ell)})^2 Y_{\ell m}}
=\ell ~(\ell+1)~\widetilde{  Y_{\ell m}}.
\end{equation*}  
We recall that the $(\widetilde{ Y_{\ell m}})_{\ell \leqslant 2j}$ form a 
basis of ${\cal O} ^{\sigma j}$.
The relations above make $\widetilde{ Y_{\ell m}}$ 
appear as the unique (up to a constant) element  of   ${\cal O} ^{\sigma j}$ that is common eigenvector to $ \calL ^{\sigma j}_3$ and $ (\calL ^{\sigma j})^2$, with eigenvalues $m$ and $\ell ~(\ell +1)$ respectively.
This implies  by linearity  that for all $f$ such that $\widetilde{f}$ makes sense
\begin{equation*}
\calL ^{\sigma j} _a \widetilde{ f}  =\widetilde{   J_a f}~\mbox{and}~ (\calL ^{\sigma j}) ^2  \widetilde{f}   =\widetilde{J^2 f}.
\end{equation*}     

%%%%%%%%%%%%%%%%%%%%%SECTION4%%%%%%%%%%%%%%%%%%%%%
 
\section{Link with the fuzzy sphere}
\subsection{The construction of  the fuzzy sphere}

Let us  first recall an usual construction of the fuzzy sphere (see for instance 
\cite{Madore} 
p.148), that we slightly modify to make the correspondence with the CS quantization.
It starts from the decomposition of any smooth function $f \in
C^{\infty}(S^2)$ in spherical harmonics,
\begin{equation} \label{develop}
f=\sum _{{\ell}=0}^{\infty}~\sum _{m=-\ell}^{\ell}~f_{{\ell}m}~Y_{\ell m}.
\end{equation} 
Let us denote by $V^{\ell }$  the $(2 \ell +1)$-dimensional
vector space generated by the $Y_{\ell m}$, at fixed~$\ell$. 

Through the embedding of $S^2$ in $\setR^3$, any function in
$S^2$ can be seen as the restriction of a  function
on~$\setR^3$ (that we write with the same notation), and under some mild conditions  such functions 
are generated by the homogeneous polynomials in $\setR^3$. This allows us to 
express (\ref{develop})
in a polynomial form in $\setR^3$:
\begin{equation}
\label{developPol}
f(x)=f_{(0)}+\sum _{(i_1)}~f_{(i)}~x^i+...+\sum _{(i_1 i_2 ...
i_{\ell})} ~
f_{(i_1i_2...i_{\ell})}~x^{i_1}~x^{i_2}...x^{i_{\ell}}+...,
\end{equation}
where each sum subtends a $V^{\ell}$ and    involves   all 
symmetric combinations of the~ $i_k$ indices, each varying from 1 to 3. This 
gives, for each fixed value of $\ell$,
$2{\ell}+1$ coefficients $f_{(i_1i_2...i_{\ell})}$ ($\ell$ fixed), which  are
those of a symmetric traceless
$3 \times 3 \times ....\times 3$ (${\ell}$ times) tensor.

The fuzzy sphere  with $2j+1 $ cells  is usually written $S_{\rm 
fuzzy,j}$, with $j$ an integer or semi-integer.  Here, our slightly modified procedure leads to a different fuzzy sphere that we write  $ _\sigma S_{\rm 
fuzzy,j}$. We detail the steps of its standard definition.  
\begin{enumerate}
\item We consider a
 $2j+1$ dimensional irreducible unitary
representation  (UIR ) of SU(2). The standard construction  considers
the  vector space $V^j$ of dimension 
 $2j+1$, on which  the 
three generators  of SU(2) are  expressed as the usual $(2j+1) \times (2j+1) $ Hermitian matrices $J_a$. 
Here we will make a different choice, namely the three SAMOs $\Lambda _j$, which correspond to the choice of the representation space $ \calH ^{\sigma j}$ (instead of 
$V^j$ in the usual construction). Since they
 obey the commutation relations of SU(2),
\begin{equation}\label{COMM}  
[\Lambda ^{\sigma j}_a,\Lambda ^{\sigma j}_b]=i~\epsilon _{abc}~\Lambda ^{\sigma j}_c,
\end{equation} the usual procedure may be applied. 
As we have seen,  $ \calH ^{\sigma j}$ can be realized as the Hilbert space spanned by the spin spherical harmonics $\{_\sigma Y_{j\mu}\}_{\mu=-j...j}$, with the usual inner product. The 
latter provide the SSH (ON) basis.

Since   the standard derivation of all properties of the fuzzy sphere rest only upon the abstract commutation rules (\ref{COMM}), nothing but the representation space changes if we  adopt the representation space $\calH$ instead of $V$. 

\item The operators $\Lambda_a^{\sigma j}$ belong  to $ \calO ^{\sigma j}$, and have a Lie algebra structure, through the skew products defined by the  commutators. But the \emph{symmetrized} products of operators provide   a second algebra structure, that we write $\calO ^{\sigma j}$,  at the basis of the construction of the fuzzy sphere: these symmetrized  products of 
the $\Lambda_a^{\sigma j}$, up to power~$2j$, generate the algebra
$\calO ^{\sigma j}$
  (of dimension $(2j+1)^2$)  of all  linear endomorphisms 
of $\calH ^{\sigma j},$ exactly like the ordinary $J_a$ do in the original Madore construction.
This is  the standard construction of the fuzzy sphere, with the $J_a$ and $V^j$ replaced by $\Lambda ^{\sigma j}_a$ and $\calH ^{\sigma j}$.

\item The construction of the fuzzy sphere (of radius $r$) is defined by  associating 
 an  operator $\hat{f}$ in $ \calO ^{\sigma j} $  to any function $f$. Explicitly, this is done by first
replacing each coordinate  $x^i$ by the operator 
\begin{equation}
\label{ansatzmad}
\widehat{x ^a} \equiv\kappa~\Lambda ^{\sigma j} _a \equiv  \frac{ r~\Lambda ^{\sigma j}_a}
{\sqrt{ j( j+1)}},
\end{equation}
 
in the above expansion (\ref{developPol}) of $f$ (in the usual construction, this would be $J_a$ instead of $\Lambda ^{\sigma j}_a$). Next, we replace in (\ref{developPol})   the usual product by the  
symmetrized product of operators,  and  we truncate the sum at index $\ell=2j$. This 
associates to any  function $f$ an  operator $\hat{f} \in   \calO ^{\sigma j} $.

\item The   vector space  $\mbox{Mat}_{2j+1}$ 
of   $(2j+1) \times (2j+1) $ matrices  is  linearly generated by a number 
$(2j+1)^2$ of independent  matrices.   According to the above  construction, a basis of $\mbox{Mat}_{2j+1}$    can be taken as all the products of the $\Lambda _a^{\sigma j}$ up to power~$2j+1$ (which is 
necessary and sufficient to close the algebra).

\item The commutative algebra limit is restored by letting $j$ go to the infinity while parameter $\kappa$ goes to zero and $\kappa j$ is fixed to $\kappa j = r$.
\end{enumerate} 

The geometry of the fuzzy sphere $S_{\rm fuzzy,j}$ is thus constructed  after  making
the choice of the algebra of the matrices of the 
representation, with their matrix product. It is taken as the algebra of 
 operators, which generalize the  functions. The rank $(2j+1)$ of the
matrices invites us to view them as acting as  endomorphisms in an
Hilbert space of dimension $(2j+1)$. This is exactly what allows the
coherent states quantization introduced in the previous section.

\subsection{Operators}

We have defined the action  on $ \calO ^{\sigma j}$:
$$ { \calL ^{\sigma j} _a} A\equiv [\Lambda ^{\sigma j}_a, A].$$

The formula (\ref{developPol}) expresses any   function $f$ of $V^{\ell}$ as the 
reduction to $S^2$ of an  homogeneous polynomials  homogeneous of order $\ell$:
$$f=\sum_{\alpha,\beta,\gamma}~f_{\alpha,\beta,
\gamma}~(x^1)^\alpha~(x^2)^\beta~(x^3)^\gamma;~ \alpha+\beta+\gamma=\ell.$$

The action of the  ordinary momentum operators $ J _3$ and $J^2$ is straightforward. Namely,
$$
J_3f=\sum_{\alpha,\beta,\gamma}~f_{\alpha,\beta,\gamma}~(-i)~\left\lbrack\beta 
(x^1)^{\alpha+1}~(x^2)^{\beta-1}~(x^3)^\gamma-\alpha (x^1)^{\alpha-1}~(x^2)^{\beta+1}~(x^3)^\gamma\right\rbrack,
$$
and similarly for $J_1$ and $J_2$.

On the other hand, we have by definition
\begin{equation}
\label{f}
 \hat{f}=\sum_{\alpha,\beta,\gamma}~f_{\alpha,\beta,
\gamma}~S\left( (\widehat{x^1}) ^\alpha ~( \widehat{x^2}) ^\beta ~(\widehat{x^3} )^\gamma  \right),\end{equation}
where $S(\cdot)$  means symmetrization.
Recalling $\widehat{x ^a}=\kappa ~\Lambda ^{\sigma j} _a$, and using (\ref{COMM}), we apply the 
operator $ { \calL ^{\sigma j} _3}$ to this expression:\begin{equation}
\label{A1}
 { \calL ^{\sigma j}  _3}\hat{f}\equiv [\Lambda ^{\sigma j} _3,\hat{f}] =\sum_{\alpha,\beta,\gamma}~f_{\alpha,\beta,
\gamma}~\left\lbrack\Lambda ^{\sigma j} _3,S \left(\hat{x^1}^\alpha~\hat{x^2}^\beta~\hat{x^3}^\gamma \right)\right\rbrack.
\end{equation}

We prove in appendix  B that the commutator of the symmetrized is the symmetrized  of the commutator.
 Then, using  the identity 
$$[J,AB\cdots M]= [J,A]~B \cdots M+ A~[J,B] \cdots M+\cdots +AB\cdots [J,M],$$
 which results easily (by induction)  
 from $ [J,AB] = [J,A]~B +A~[J, B] $, it follows that
\begin{equation}
\label{Proof1}
  \calL ^{\sigma j} _3 \hat{f}\equiv [\Lambda ^{\sigma j} _3,\hat{f}]=\sum_{\alpha,\beta,\gamma}~f_{\alpha,
\beta,\gamma}~\left(
i \alpha~
\hat{x^1}^{\alpha-1}~\hat{x^2}^{\beta+1}~\hat{x^3}^\gamma -i\beta~
\hat{x^1}^{\alpha+1}~\hat{x^2}^{\beta-1}~\hat{x^3}^\gamma \right).
\end{equation}
 
We thus have  proven 
$$ { \calL ^{\sigma j} _3}\hat{f}=\widehat{J_3 f}.$$
Similar identities hold for $ { \calL ^{\sigma j} _1},~ { \calL ^{\sigma j} _2}$ and thus for  $ ({ \calL }^{\sigma j} )^2$. 

It results that    $\widehat{ Y_{\ell m}}$ appears as  an  element of $\calO ^{\sigma j}$ which is a common eigenvector 
of ${\calL} ^{\sigma j} _3$, with value $m$, and of $({\calL}^{\sigma j} )^2$, with value $\ell (\ell+1)$.
Since we have proved above that such an element is   unique (up to a constant), it results that each 
$\widehat{Y_{\ell m}}\propto  \widetilde{Y_{\ell m}}$. Thus, 
 the $\widehat{Y_{\ell m}}$'s, for $\ell \le j$, $-j\le m\le j$ form a basis of 
${\cal A}^j$.

Then,  the {Wigner-Eckart} theorem (see \ref{Wigner-Eckart}) implies that
$\widetilde{Y_{\ell m}} = C(\ell)~ \widehat{Y_{\ell m}}$, where the proportionality constant $C(\ell)$ does not depend on $m$ (what can also be checked directly).

These coefficients can be calculated directly, after remarking that 
$$\widehat{Y_{\ell \ell}}\propto (\Lambda_+)^\ell
\propto 
(\widehat{x^1}+i~Ê\widehat{x^2})^\ell.$$

In fact, $$\widehat{Y_{\ell \ell}}= a(\ell) ~Ê  
(\widehat{x^1}+i~Ê\widehat{x^2})^\ell; \ a(\ell) ~Ê =\frac{\sqrt{(2\ell +1)!}}{2^{\ell+1}~Ê\sqrt{\pi}~ \ell !}.$$

We obtain 

$$C(\ell)
= 2^{\ell } 
~\frac{ (-1)^{j+\sigma -2~\ell}~(2~Êj+1)  }{\kappa^\ell ~
 }~Ê\sqrt{
\frac{ Ê(2j-\ell)! ~Ê   }{ (2j+\ell +1)! Ê} } ~\begin{pmatrix}
 j  & j & \ell  \\
-\sigma  & \sigma & 0
\end{pmatrix}.$$

%%%%%%%%%%%%%%%%%%%%%CONCLUSION%%%%%%%%%%%%%%%%%%%%%
\section{Discussion}\label{discussion}

\begin{table}
  \centering 
 \begin{tabular}{|c|c|c|}
\hline
&coherent states&Madore-like\\
&fuzzy sphere&fuzzy sphere\\
\hline
\hline
Hilbert space&\multicolumn{2}{|c|}{${\cal H} ={\cal H}^{\sigma j}=\mbox{span} (_\sigma Y_{j \mu}) \subset L^2(S^2)$}\\
\hline
&\multicolumn{2}{|c|}{ } \\
endomorphisms &\multicolumn{2}{|c|}{${\cal O}={\cal O}^{\sigma j} =\mbox{End}  { \cal H} ^{\sigma j} $} \\
\hline
spin angular momentum&\multicolumn{2}{|c|}{ } \\
 operators&\multicolumn{2}{|c|}{$\Lambda ^{\sigma j}_a \in {\cal O}  $} \\
\hline
 && \\ 
observables &$\widetilde{f}\in {\cal O} ^{\sigma j};~ \widetilde{x^a} =K ~\Lambda ^{\sigma j} _a $  & $\widehat{f}\in {\cal O} ^{\sigma j};~\widehat{x^a} =\kappa ~\Lambda ^{\sigma j} _a $\\
 &&\\
\hline 
  &&\\
action of angular momentum &$\calL ^{\sigma j} _a  \widetilde{f}\equiv[ \Lambda ^{\sigma j} _a,  \widetilde{f}] =   \widetilde{J_a~f}$&
$\calL ^{\sigma j} _a  \widehat{f}\equiv[ \Lambda ^{\sigma j} _a,   \widehat{f}]  =   \widehat{J_a~f}$\\
  &&\\
  \hline
  &\multicolumn{2}{|c|}{ } \\
correspondence&\multicolumn{2}{|c|}{$ \widetilde{Y_{\ell m}}=C(\ell)~\widehat{Y_{\ell m}}$}\\
\hline 
\hline
\end{tabular}
 
  \caption{Coherent state quantization of the sphere is compared to the standard construction of the fuzzy sphere through correspondence formula.}\label{table1}
\end{table}

We thus have  two families of quantization of the sphere.
\begin{itemize}
  \item The usual construction of the fuzzy sphere, which depends  on the parameter $j$. This parameter  defines the \guill size" of the discrete cell.
  \item  The present construction coherent states which makes use of coherent states and  which depends on two parameters, $j$ and $\sigma \ne 0$.
\end{itemize}
These two quantizations may be formulated as involving the same algebra of operators (quantum observables) $\calO$, acting on the same Hilbert space $\calH$ (see Table \ref{table1}). Note that $\calH$  and $\calO$ are not the Hilbert space and  algebra   usually involved in the usual expression of the fuzzy sphere (when we consider them as embedded in the space of  functions of the spheres, and of operators acting on them), but they are isomorphic to them, and nothing is changed.

The difference lies in the fact that the quantum counterparts, $\tilde{f}$ and $\hat{f}$  of a given classical observable $f$ differ in both approaches. Thus, the CS quantization really differs from the usual fuzzy sphere quantization. This raises the question iof whether  the CS quantization is or is not a construction of a new type ofÊfuzzy sphere. It results from the calculations above that all properties of the usual fuzzy sphere are shared by the CS quantized version. The only point to be checked is if it gives the sphere manifold in some classical limit. The answer is positive as far as the classical limit is correctly defined. Simple calculations show that it is obtained as the limit $j \mapsto \infty,~\sigma \mapsto \infty$, provided that the ratio $\sigma/j$ tends to a finite value. Thus, one may consider that the CS quantization leads to a one parameter family of fuzzy spheres if we impose relations of the type $\sigma =j - \sigma_0$, for fixed $\sigma_0> 0 $  (for instance).

%%%%%%%%%%%%%%%%%%%%%CONCLUSION%%%%%%%%%%%%%%%%%%%%%
\section{Conclusion}

We have proposed a  general quantization procedure which applies to   any measurable  set $X$. It  proceeds 
from the choice of    an Hilbert space $\calH$ of prescribed dimension. We 
have presented in details an implementation of this procedure    (non 
necessarily unique)  from   an explicit 
family of coherent states, which realizes a natural embedding of $X$ 
into $\calH$.

We have  applied   this    CS procedure to the sphere 
$S^2$. We  started  from a natural 
basis linked to the UIR's of   the group SU(2): for any value of $j$ and $\sigma $, we 
chose the Hilbert space  $\calH^{\sigma j}$, which carries a  UIR 
of SU(2). 
Our  CS construction  associates, to any classical observable $f \in L^2$, a quantum   observable $\widetilde{f}$, which belong to the 
algebra of endomorphisms $\calO^{\sigma j}\equiv \End (\calH^{\sigma j})$. On the other hand, we also followed the usual   fuzzy sphere construction (with 
$2j+1$ cells), by replacing the coordinates by operators acting on the same Hilbert space.
This  allowed us  to associate a fuzzy observable $\widehat{f}$ to any classical observable $f$.
Those form     the algebra of operators acting on the fuzzy sphere. 

For the particular classical observables provided by the ordinary spherical harmonics,
we have shown that the CS quantum observable and the fuzzy observable  
 coincide up to a constant, $\widehat{Y_{\ell m}}=C(\ell)~ \widetilde{Y_{\ell m}}$, and the explicit value of this constant has been given. However, in general, $\widetilde{f}$ differs from $\widehat{f}$, although the correspondence is easy established from the relation above, through a development in the usual spherical harmonics.
 
Thus, the CS quantization procedure really differs from the construction of the usual fuzzy sphere. Although they share  the same algebra of quantum observables, acting on the same Hilbert space, the CS quantum observables $\widetilde{f}$ and the fuzzy one, $\widehat{f}$, associated to the same  classical observable $f$ differ. And there is no way to make them coincide, since the CS quantization with $\sigma=0$  leads to trivial results. 

Our discussion in (\ref{discussion}) allows us    to consider our  CS quantization procedure     as a      construction of a new type of fuzzy sphere, with properties differing from the standard one. It shares most of the properties of the usual fuzzy sphere, but appears more economic in the sense that \\
- it does not require a group action on the space to be quantized;\\
- it does not require an initial expansion of the functions into spherical harmonics.

Applications  of procedures of this type to the sphere have appeared in different 
contexts. 
For instance,     a similar procedure is carried out in  \cite{Taylor} in order to achieve a  
\emph{regularization} of a membrane, with surface $S^2$, by a 
mapping of functions to matrices, similar to the one presented here. 
Despite analog mathematics, the procedure there is not seen as   a 
\emph{quantization} and, according to the author,  the regularized theory 
still   requires a further quantization. Similar regularization exists 
for surfaces of arbitrary genius, and it would be  interesting to 
apply the CS procedure in these cases. Also, it should not be  difficult to 
explore cases with more dimensions, and in particular  $S^3$. This offers 
possibilities to construct new fuzzy versions of these spaces. 
Moreover, authors in \cite{Freidel} have given a description of the 
fuzzy sphere in terms of SU(2) spin networks. Since the latter play an 
important role in the canonical quantization of \gr, this suggests that the 
application of the CS procedure to the quantization of gravity or to various geometries, compact or non-compact \cite{gamouque} could be 
fruitful, a program that we start to explore. Furthermore, 
the universality of the CS procedure would allow explicit constructions  of    
spin networks associated to  different groups, in particular SU(3). Since it has claimed 
that the latter could be of 
importance for quantum gravity, this reveals to be a promising   field of research  
also.

%%%%%%%%%%%%%%%%%%%%APPENDICES%%%%%%%%%%%%%%%%%%%%%%%%

\section{Appendix A: Spin spherical harmonics}
 %%%%%%%%%%%%%%%%%%%%%%%%%%%%%%%%%%%%%%%%%%%%%%
\subsection{$SU(2)$-parameterization}
\begin{equation}
SU(2) \ni \xi = \left(\begin{array}{cc}\xi_0 + i \xi_3 & -\xi_2 + i \xi_1 \\ \xi_2 + i \xi_1 & \xi_0 - i\xi_3\end{array}\right).
\end{equation}
In bicomplex angular coordinates,
\begin{align}
\label{angsu2}
 \xi_0 + i \xi_3 = \cos{\omega} e^{i\psi_1},  & \ \ 
  \xi_1 + i \xi_2 = \sin{\omega} e^{i\psi_2} \\
  0 \leqslant \omega \leqslant \frac{\pi}{2}, & \ \  0 \leqslant \psi_1, \psi_2  < 2 \pi.
\end{align}
and so 
\begin{equation}
SU(2) \ni \xi = \left(\begin{array}{cc} \cos{\omega} e^{i \psi_1}&  i\sin{\omega} e^{i\psi_2} \\  i\sin{\omega} e^{-i\psi_2} & \cos{\omega} e^{-i\psi_1}\end{array}\right),
\end{equation}
in agreement with Talman \cite{Talman}.
\subsection{Matrix elements of $SU(2)$-UIR}
\begin{align}
\label{matelsu2}
\nonumber D^j_{m_1 m_2} (\xi) &= (-1)^{m_1 - m_2} \left\lbrack (j+m_1)! (j-m_1)! (j+m_2)! (j-m_2)! \right\rbrack^{1/2} \times \\
\times& \sum_{t} \frac{( \xi_0 + i \xi_3)^{j-m_2 - t}}{(j - m_2 -t)!} \, \frac{( \xi_0 - i \xi_3)^{j +m_1 - t}}{(j +  m_1 -t)!}\,\frac{(- \xi_2 + i \xi_1)^{t+m_2 - m_1}}{(t +m_2 - m_1)!}  \frac{( \xi_2+ i \xi_1)^{ t}}{t!} \, ,
\end{align}
in agreement with Talman.
With angular parameters
  the matrix elements of the UIR of $SU(2)$ are given in terms of Jacobi polynomials \cite{magnus-ob} by:
\begin{align}
\label{jacDj}
    \nonumber D^j_{m_1 m_2} (\xi) &=e^{-im_1(\psi_1 + \psi_2)} e^{-im_2(\psi_1 - \psi_2)} i^{m_2 - m_1} \sqrt{\frac{(j-m_1)!(j+m_1)!}{(j-m_2)!(j+m_2)!}}\times \\
\times&  \frac{1}{2^{m_1}}\, (1 + \cos{2\omega})^{\frac{m_1 + m_2}{2}}\, (1 - \cos{2\omega})^{\frac{m_1 - m_2}{2}}  P_{j-m_1}^{(m_1 - m_2,m_1 + m_2)}(\cos{2\omega}), 
\end{align}
in agreement with Edmonds \cite{Edmonds} (up to an irrelevant phase factor).

%%%%%%%%%%%%%%%%%%%%%%%%%%%%%%%%%%%%%%%%%%%%%%

\subsection{Orthogonality relations and $3j$-symbols}
Let us equip the $SU(2)$ group with its Haar measure :
\begin{equation}
\label{haar}
\mu(d\xi) = \sin{2\omega}\, d\omega \, d\psi_1 \, d\psi_2,
\end{equation}
in terms of the bicomplex angular parametrization.
Note that the volume of $SU(2)$ with this choice of normalization is $8\pi^2$.
The orthogonality relations  satisfied by the matrix elements $D^j_{m_1 m_2} (\xi)$ reads as:
\begin{equation}
\label{orthogrel}
\int_{SU(2)} D^j_{m_1 m_2} (\xi)\left(D^{j'}_{m'_1 m'_2} (\xi) \right)^{\ast}\, \mu(d\xi) = \frac{8 \pi^2}{2j+1}\, \delta_{jj'}  \delta_{m_1m'_1} \delta_{m_2m'_2}.
\end{equation}
in connection with the reduction of the tensor product of two UIR's of $SU(2)$, we have the following equivalent formula involving the so-called $3-j$ symbols (proportional to Clebsch-Gordan coefficients), in the Talman notations :
\begin{equation}
\label{CGdev}
D^j_{m_1 m_2} (\xi)\,D^{j'}_{m'_1 m'_2} (\xi) = \sum_{j'' m''_1 m''_2}(2j'' + 1) \begin{pmatrix}
   j   &  j' & j''  \\
   m_1   &  m'_1 & m''_1
\end{pmatrix} \begin{pmatrix}
   j   &  j' & j''  \\
   m_2   &  m'_2 & m''_2
\end{pmatrix} \, \left(D^{j''}_{m''_1 m''_2} (\xi)\right)^{\ast},
\end{equation}
\begin{equation}
\label{CGint}
\int_{SU(2)} D^j_{m_1 m_2} (\xi)\,D^{j'}_{m'_1 m'_2} (\xi)\, D^{j''}_{m''_1 m''_2} (\xi)\, \mu(d\xi) = 8 \pi^2 \,   \begin{pmatrix}
   j   &  j' & j''  \\
   m_1   &  m'_1 & m''_1
\end{pmatrix} \begin{pmatrix}
   j   &  j' & j''  \\
   m_2   &  m'_2 & m''_2
\end{pmatrix} .
\end{equation}
One of the multiple expressions of the $3-j$ symbols (in the convention that there are all real) is given by:
\begin{align}
\label{3j}
\nonumber \begin{pmatrix}
   j   &  j' & j''  \\
   m   &  m' & m''
\end{pmatrix} =& (-1)^{j-j'-m''}\left\lbrack\frac{(j +j'-j'')! (j -j'+j'')!(-j +j' +j'')!}{(j +j' +j'' +1)!} \right\rbrack^{1/2} \\
\times \sum_s (-1)^s & \frac{\left\lbrack (j+m)! (j-m)!(j'+m')! (j'-m')!(j''+m'')! (j''-m'')!\right\rbrack^{1/2}}{s! (j'+m' -s)!(j-m-s)!(j''-j'+m +s)!(j'' -j -m' +s)! (j + j' -j''-s)!}
\end{align}

%%%%%%%%%%%%%%%%%%%%%%%%%%%%%%%%%%%%%%%%%%%%%

\subsection{Spin spherical harmonics}

The spin spherical harmonics, as functions on the 2-sphere $S^2$ are defined as follows:
\begin{align}
\label{spinsphar}
_{\sigma}Y_{j \mu}(\hat{\mathbf{r}}) &= \sqrt{\frac{2j + 1}{4\pi}} \left\lbrack D^j_{\mu \sigma} \left(\xi\left( \mathcal{R}_{\hat{\mathbf{r}}}\right)\right)\right\rbrack^{\ast} = (-1)^{\mu-\sigma}\sqrt{\frac{2j + 1}{4\pi}} D^j_{-\mu -\sigma} \left(\xi\left( \mathcal{R}_{\hat{\mathbf{r}}}\right)\right)\\
& = \sqrt{\frac{2j + 1}{4\pi}} D^j_{ \sigma \mu}  \left(\xi^{\dag}\left( \mathcal{R}_{\hat{\mathbf{r}}}\right)\right),
\end{align}
where $\xi\left( \mathcal{R}_{\hat{\mathbf{r}}}\right)$ is a (nonunique) element of $SU(2)$ which corresponds to the space rotation $\mathcal{R}_{\hat{\mathbf{r}}}$ which brings the unit vector $\widehat{\mathbf{ e_3}}$ to the unit vector $\widehat{\mathbf{r}}$ with polar coordinates :
\begin{equation}
\label{polcoor}
\widehat{\mathbf{r}} = \left\lbrace \begin{array}{ll}
x^1 =    & \sin\theta \cos\phi ,  \\
x^2 =    & \sin\theta \sin\phi ,  \\
x^3 =    & \cos\theta .     
\end{array} \right.
\end{equation}
We immediately infer from the definition (\ref{spinsphar}) the following properties:
\begin{equation}
\label{inver}
\left( _{\sigma}Y_{j \mu}(\hat{\mathbf{r}})  \right)^{\star} = (-1)^{\sigma - \mu} \, {}_{-\sigma}Y_{j -\mu}(\hat{\mathbf{r}}), 
\end{equation}
\begin{equation}
\label{sumharmsq}
\sum_{\mu = -j}^{\mu = j} \left\vert _{\sigma}Y_{j \mu}(\hat{\mathbf{r}}) \right\vert^2 = \frac{2j + 1}{4\pi}.
\end{equation}
Let us recall here the correspondence (homomorphism) $\xi = \xi(\mathcal{R}) \in SU(2) \leftrightarrow \mathcal{R} \in S0(3) \simeq SU(2) / \Z_2 $ :
\begin{align}
\label{su2so3}
\widehat{\mathbf{r}}' &= (x'_1,x'_2,x'_3) =  \mathcal{R}\cdot \widehat{\mathbf{r}} \longleftrightarrow       \\
  \left(\begin{array}{cc} i x'_3 & -x'_2 + i x'_1 \\ x'_2 + i x'_1 &  - i x'_3\end{array}\right)   &=
  \xi \left(\begin{array}{cc} i x_3 & -x_2 + i x_1 \\ x_2 + i x_1 &  - i x_3\end{array}\right) \xi^{\dag}.
\end{align}
In the particular case of (\ref{spinsphar}) the angular coordinates $\omega, \psi_1, \psi_2$ of the $SU(2)$-element $\xi\left( \mathcal{R}_{\hat{\mathbf{r}}}\right)$ are constrained by
\begin{align}
\label{const}
 \cos{2\omega} =& \cos{\theta}, \ \sin{2\omega} = \sin{\theta}, \qquad \mathrm{so} \qquad 2 \omega = \theta, \\
e^{i(\psi_1 + \psi_2)} =& i e^{i \phi} \qquad \mathrm{so} \qquad 
  \psi_1 + \psi_2  = \phi + \frac{\pi}{2}.
\end{align}
Here we should pay a special attention to the range of values for the angle $\phi$, depending on whether $j$ and consequently $\sigma$ and $m$ are half-integer or not. If $j$  is half-integer, then angle $\phi$ should be defined $\mod(4\pi)$ whereas if $j$ is integer, it should be defined  $\mod(2\pi)$.

We still have one degree of freedom concerning the pair of angles $\psi_1$, $\psi_2$. We leave open the option concerning the $\sigma$-dependent phase factor by putting
\begin{equation}
\label{sigphase}
i^{-\sigma} e^{i \sigma (\psi_1 - \psi_2)} \stackrel{\mathrm{def}}{=} e^{i \sigma \psi},
\end{equation} 
where $\psi$ is arbitrary.
With this choice and considering (\ref{matelsu2}) we get the expression of the spin spherical harmonics in terms of $\phi$, $\theta/2$ and $\psi$: 
\begin{align}
\label{spinpol1}
\nonumber  _{\sigma}Y_{j \mu}(\hat{\mathbf{r}}) &= (-1)^{\sigma} e^{i \sigma \psi} e^{i \mu \phi}\sqrt{\frac{2j + 1}{4\pi}} \sqrt{\frac{ (j+ \mu)! (j-\mu)!}{(j+\sigma)! (j-\sigma)!}} \times \\
\times& \left(\cos{\frac{\theta}{2}}\right)^{2j} \sum_{t} (-1)^t \begin{pmatrix}
j-\sigma     \\
 t     
\end{pmatrix} \begin{pmatrix}
j + \sigma         \\
  t + \sigma - \mu 
\end{pmatrix}  \left(\tan{\frac{\theta}{2}}\right)^{2t + \sigma - \mu} , \\
\nonumber &= (-1)^{\sigma} e^{i \sigma \psi} e^{i \mu \phi}\sqrt{\frac{2j + 1}{4\pi}} \sqrt{\frac{ (j+\mu)! (j-\mu)!}{(j+\sigma)! (j-\sigma)!}} \times \\
\times& \left(\sin{\frac{\theta}{2}}\right)^{2j} \sum_{t} (-1)^{j -t + \mu - \sigma} \begin{pmatrix}
j-\sigma     \\
 t - \mu 
\end{pmatrix} \begin{pmatrix}
j + \sigma         \\
  t + \sigma       
\end{pmatrix}  \left(\cot{\frac{\theta}{2}}\right)^{2t + \sigma - \mu} ,  
      \end{align}
      which are  not in agreement with the definitions of  Newman and Penrose \cite{Newman}, Campbell \cite{Campbell} (note there is a mistake in the expression given by Campbell, in which a $\cos{\frac{\theta}{2}}$ should read $\cot{\frac{\theta}{2}}$), and Hu and White \cite{HuWhite}. Besides presence of different phase factors, the disagreement is certainly due to a different relation between the polar angle $\theta$ and the Euler angle.

Now, considering (\ref{jacDj}), we get the expression of the spin spherical harmonics in terms of the Jacobi polynomials, valid in the case in which $ \mu \pm \sigma > -1$:
\begin{align}
\label{jacspin}
    \nonumber _{\sigma}Y_{j \mu}(\hat{\mathbf{r}}) &=  (-1)^\mu e^{i \sigma \psi}\sqrt{\frac{2j + 1}{4\pi}}  \sqrt{\frac{(j-\mu)!(j+\mu)!}{(j-\sigma)!(j+\sigma)!}}\times \\
\times&  \frac{1}{2^{\mu}}\, (1 + \cos{\theta})^{\frac{\mu + \sigma}{2}}\, (1 - \cos{\theta})^{\frac{\mu - \sigma}{2}}  P_{j-\mu}^{(\mu - \sigma,\mu + \sigma)}(\cos{\theta}) \,e^{i \mu \phi}.
\end{align}
For other cases, it is necessary to use alternate expressions based on the relations \cite{magnus-ob}:
\begin{equation}
\label{SpecJac}
P^{(-l, \beta)}_{n} (x) = \frac{\binom{n + \beta}{l}}{\binom{n }{l}}\left(\frac{x-1}{2}\right)^l P^{(l, \beta)}_{n-l} (x), \ P^{(\alpha, \beta)}_{0} (x) = 1.
\end{equation}

Note that with $\sigma = 0$ we recover the expression of the normalized spherical harmonics :
\begin{align}
\label{spharm}
\nonumber _{0}Y_{jm}(\hat{\mathbf{r}}) &= Y_{jm}(\hat{\mathbf{r}}) =  (-1)^{m} \sqrt{\frac{2j + 1}{4\pi}}  \sqrt{(j-m)!(j+m)!}
  \frac{1}{j!\, 2^{m}}\, (\sin{\theta})^m  P_{j-m}^{(m ,m )}(\cos{\theta}) \,e^{im\phi} \\
  & =    \sqrt{\frac{2j + 1}{4\pi}}\, \sqrt{\frac{(j-m)!}{(j+m)!}}\, P_j^m (\cos{\theta}) e^{i m\phi}
\end{align}
since we have  the following relation between associated Legendre polynomials and Jacobi polynomials
\begin{equation}
\label{asslegjac}
P_{j-m}^{(m,m)}(z) = (-1)^m 2^m (1- z^2)^{-\frac{m}{2}}  \frac{j!}{(j+m)!} P_j^m(z),
\end{equation}
for $m>0$. We recall also  the symmetry formula
\begin{equation}
\label{symass}
P_j^{-m} (z) = (-1)^m \frac{(j-m)!}{(j+m)!} P_j^m(z).
\end{equation}
Our expression of spherical harmonics is rather standard, in agreement with Arkfen \cite{Arkfen,Weisstein}\footnote{Sometimes (e.g., Arfken 1985 \cite{Arkfen}), the Condon-Shortley phase $(-1)^m$ is prepended to the definition of the spherical harmonics. Talman adopted this convention.}

%%%%%%%%%%%%%%%%%%%%%%%%%%%%%%%%%%%%%%%%%%%%%%%

\subsection{Transformation laws}
We consider here the transformation law of the spin spherical harmonics under the rotation group. From the relation 
\begin{equation}
\label{ rotrot}
\mathcal{R}  \mathcal{R}_{^t\mathcal{R}\hat{\mathbf{r}}} =  \mathcal{R}_{\hat{\mathbf{r}}}
\end{equation}
for any $\mathcal{R} \in SO(3)$, and from the homomorphism $\xi(\mathcal{R}\mathcal{R}') = \xi(\mathcal{R})\xi(\mathcal{R}')$ between $SO(3)$ and $SU(2)$, we deduce from the definition (\ref{spinsphar}) of the spin spherical harmonics the transformation law
\begin{align}
\label{transrot}
\nonumber  _{\sigma}Y_{j \mu}(^t\mathcal{R}\cdot \hat{\mathbf{r}}) &= \sqrt{\frac{2j + 1}{4\pi}} D^j_{ \sigma \mu}  \left(\xi^{\dag}\left( \mathcal{R}_{^t\mathcal{R}\cdot\hat{\mathbf{r}}}\right)\right) = 
 \sqrt{\frac{2j + 1}{4\pi}} D^j_{ \sigma \mu}  \left(\xi^{\dag}\left( ^t\mathcal{R}\mathcal{R}_{\hat{\mathbf{r}}}\right)\right)\\
\nonumber & = \sqrt{\frac{2j + 1}{4\pi}} D^j_{ \sigma \mu}  \left(\xi^{\dag}\left( \mathcal{R}_{\hat{\mathbf{r}}}\right)\xi\left( \mathcal{R}\right)\right)  =  \sqrt{\frac{2j + 1}{4\pi}} \sum_{\nu}D^j_{ \sigma \nu}  \left(\xi^{\dag}\left( \mathcal{R}_{\hat{\mathbf{r}}}\right)\right) D^j_{ \nu \mu}  \left(\xi\left( \mathcal{R}\right)\right)\\
 & =  \sum_{\nu} \, _{\sigma}Y_{j\nu} ( \hat{\mathbf{r}})~ D^j_{ \nu \mu}  \left(\xi\left( \mathcal{R}\right)\right) ,
\end{align}
as expected if we think to the special case ($\sigma = 0$) of the spherical harmonics.

Given a function $f( x)$ on the sphere $S^2$ belonging to the $2j+1$-dimensional Hilbert space $\calH ^{\sigma j}$
and a rotation $\mathcal{R} \in SO(3)$, we define the rotation  operator $\mathcal{D}^{\sigma j}(\mathcal{R})$  for that representation
%,  in $\calH^{\sigma j}$, 
by
\begin{equation}
\label{reprot}
\left(\mathcal{D}^{\sigma j} (\mathcal{R})f\right)(x) = f(\mathcal{R}^{-1}\cdot x)=f({ }^{t}\mathcal{R}\cdot x).
\end{equation}
Thus, in particular,
\begin{equation}
\label{reprotY}
\left(\mathcal{D}^{\sigma j} (\mathcal{R})~ _{\sigma}Y_{j \mu} \right)(\hat{\mathbf{r}}) =~
_{\sigma}Y_{j \mu}(^t\mathcal{R}\cdot \hat{\mathbf{r}}).
\end{equation}

 The  generators of  the three   rotations $\mathcal{R}^{(a)},~a=1,2,3$, around the three usual axes, are the angular momentum operator in the representation. When $\sigma=0$, we recover the usual SHs, and these generators are the usual angular momentum operators $J^i$ (short notation for $J^{(j)}_i$) for that representation. 
In the general case $\sigma\ne 0$, we call them $\Lambda^{(\sigma j)}_a$.
%, that we abbreviate occasionally  in $\Lambda_i $. 
We study their properties below.

%%%%%%%%%%%%%%%%%%%%%%%%%%%%%%%%%%%%%%%%%%%%%%%

\subsection{Infinitesimal transformation laws}

Recalling that the components $J_a = -i~\epsilon _{abc} ~x^b~\partial _c$ of the ordinary  angular momentum operator are given in spherical coordinates by:
\begin{align}
\label{angmom}
 J_3   &= -i \partial_{\phi},   \\
 J_+ &= J_1 + iJ_2   = e^{i\phi}  \left(  \partial_{\theta} + i \cot{\theta}  ~�partial_{\phi} \right),\nonumber\\
  J_- &= J_1 - iJ_2   = -e^{-i\phi}  \left(  \partial_{\theta} - i \cot{\theta}  ~�partial_{\phi} \right)\nonumber.
\end{align}
We have  introduced the ``spin'' angular momentum operators:
\begin{align}
\label{spangmom}
 \Lambda ^{\sigma j}_3   &= J_3 = -i \partial_{\phi},   \\
\label{spangmom+} \Lambda ^{\sigma j}_+ &= \Lambda ^{\sigma j}_1 + i~\Lambda ^{\sigma j}_2   =  J_+  + \sigma \csc{\theta} ~e^{i\phi},\\
 \label{spangmom-} \Lambda ^{\sigma j}_- &= \Lambda ^{\sigma j}_1 - i~\Lambda ^{\sigma j}_2   =  J_-  + \sigma \csc{\theta} ~e^{-i\phi}.
\end{align}
They obey the expected commutation rules,
\begin{equation}
\label{comspin}
[\Lambda ^{\sigma j}_3,\Lambda ^{\sigma j}_{\pm}] = \pm \Lambda ^{\sigma j}_{\pm}, \qquad [\Lambda ^{\sigma j}_+,\Lambda ^{\sigma j}_{-}] = 2 \Lambda ^{\sigma j}_3.
\end{equation}
These operators are the  infinitesimal generators of the action of $SU(2)$ on the spin spherical harmonics:
\begin{align}
\label{angspher}
 \Lambda ^{\sigma j}_3 \, {}_{\sigma}Y_{j\mu}  &=  \mu \,\,{}_{\sigma}Y_{j\mu}\\
\label{angspher+} \Lambda^{\sigma j}_+ \, {}_{\sigma}Y_{j\mu}  &= \sqrt{(j-\mu)(j+\mu+1)}\,\, {}_{\sigma}Y_{j \mu+1}\\
 \label{angspher-} \Lambda ^{\sigma j}_- \, {}_{\sigma}Y_{j \mu} & = \sqrt{(j +\mu)(j-\mu+1)}\, \, {}_{\sigma}Y_{j \mu-1}.
\end{align}

%%%%%%%%%%%%%%%%%%%%%%%%%%%%%%%%%%%%%%%%%%%%%%%
\subsection{Integrals and $3j$-symbols}
Specifying the equation (\ref{orthogrel}) to the spin spherical harmonics lead to the following orthogonality relations which are   valid for  $j$ integer (and consequently $\sigma$ integer). 
\begin{equation}
\label{orthogspin}
\int_{S^2} \, _{\sigma}Y_{j \mu}(\hat{\mathbf{r}})\, \left(_{\sigma}Y_{j' \nu}(\hat{\mathbf{r}})\right)^{\ast} \, \mu(d\hat{\mathbf{r}})= \delta_{jj'} \delta_{\mu\nu},
\end{equation}
We recall that in the integer case, the range of values assumed by the angle $\phi$ is 
$ 0 \leqslant \phi < 2 \pi$. Now, if we consider half-integer $j$ (and consequently $\sigma$), the range of values assumed by the angle $\phi$ becomes 
$ 0 \leqslant \phi < 4 \pi$. The integral above has to be carried out on the ``doubled'' sphere $\widetilde{S}^2$ and an extra  normalization factor equal to $\dfrac{1}{\sqrt{2}}$ is needed in the expression of the spin spherical harmonics.

For a given integer $\sigma$ the set $\left\{ ~_{\sigma}Y_{j \mu}, \, -\infty \leqslant \mu \leqslant \infty, \, j \geqslant \max{(0,\sigma, m)} \right\}$ form an orthonormal basis of the Hilbert space $L^2(S^2)$. Indeed, at $\mu$ fixed  so that $\mu\pm \sigma \geqslant 0$, the set 
$$
\left\{ \sqrt{\frac{2j + 1}{4\pi}}  \sqrt{\frac{(j-\mu)!(j+\mu)!}{(j-\sigma)!(j+\sigma)!}}\, \frac{1}{2^{\mu}}\, (1 + \cos{\theta})^{\frac{\mu + \sigma}{2}}\, (1 - \cos{\theta})^{\frac{\mu - \sigma}{2}}  P_{j-\mu}^{(\mu - \sigma,\mu + \sigma)}(\cos{\theta}),\,  j\geqslant \mu \right\}
$$
is an orthonormal basis of the Hilbert space $L^2([-\pi,\pi], \sin{\theta}\, d\theta)$. The same holds for other ranges of values of $\mu$ by using  alternate expressions like (\ref{SpecJac}) for Jacobi polynomials. Then it suffices to view $L^2(S^2)$ as the tensor product  $L^2([-\pi,\pi], \sin{\theta}\, d\theta) \bigotimes L^2(S^1)$. Similar reasoning is valid for half-integer $\sigma$.
Then, the Hilbert space to be considered is the space of ``fermionic'' functions on the doubled sphere $\widetilde{S}^2$,  \emph{i.e.} such that $f(\theta, \phi + 2\pi) = - f(\theta, \phi )$.

Specifying the equation (\ref{CGdev})  to the spin spherical harmonics  leads to 
\begin{align}
\label{CGspin}
\nonumber _{\sigma}Y_{j \mu}(\hat{\mathbf{r}})\, _{\sigma'}Y_{j' \mu'}(\hat{\mathbf{r}}) &= \sum_{j'' \mu'' \sigma''}\sqrt{\frac{(2j + 1)(2j'+ 1)(2j'' + 1)}{4 \pi}} \times \\
& \times \begin{pmatrix}
   j   &  j' & j''  \\
   \mu   &  \mu' & \mu''
\end{pmatrix} \begin{pmatrix}
   j   &  j' & j''  \\
  \sigma   &  \sigma' & \sigma''
\end{pmatrix} \, \left(_{\sigma''}Y_{j'' \mu''}(\hat{\mathbf{r}})\right)^{\ast}.
\end{align}
We easily deduce from (\ref{CGspin}) the following integral involving the product of three spherical spin harmonics (in the integer case, but analog formula exists in the half-integer case) and \underline{with the constraint that} $\sigma + \sigma' + \sigma'' = 0$: 
\begin{align}
\label{CGspinint}
\nonumber \int_{S^2} \, _{\sigma}Y_{j\mu}(\hat{\mathbf{r}})\, _{\sigma'}Y_{j'\mu'}(\hat{\mathbf{r}}) \, _{\sigma''}Y_{j'' \mu''}(\hat{\mathbf{r}})\, \mu(d\hat{\mathbf{r}}) &= \sqrt{\frac{(2j + 1)(2j'+ 1)(2j'' + 1)}{4 \pi}} \times \\
& \times \begin{pmatrix}
   j   &  j' & j''  \\
 \mu   &  \mu' & \mu''
\end{pmatrix} \begin{pmatrix}
   j   &  j' & j''  \\
  \sigma   &  \sigma' & \sigma''
\end{pmatrix}.
\end{align}
Note that this formula is independent of the presence of a constant phase factor of the type $e^{i\sigma \psi}$ in the definition of the spin spherical harmonics  because of the \emph{a priori} constraint $\sigma + \sigma' + \sigma'' = 0$. On the other hand, we have to be careful in  applying Eq. (\ref{CGspinint}) because of this constraint, \emph{i.e.} since it has been derived from Eq. (\ref{CGspin}) on the ground that $\sigma''$ was already \emph{fixed} at the value $\sigma'' = -\sigma - \sigma'$. Therefore, the computation of 
$$
 \int_{S^2} \, _{\sigma}Y_{j\mu}(\hat{\mathbf{r}})\, _{\sigma'}Y_{j' \mu'}(\hat{\mathbf{r}}) \, _{\sigma''}Y_{j'' \mu''}(\hat{\mathbf{r}})\, \mu(d\hat{\mathbf{r}})
 $$
for an arbitrary triplet $(\sigma, \sigma', \sigma'')$ should be carried out independently.

%%%%%%%%%%%%%%%%%%%%%%%%%%%%%%%%%%%%%%%%%%%%%%

\subsection{Important particular case : $j = 1$}

In the particular case $j=1$, we get the following expressions for the spin spherical harmonics:
\begin{align}
 _{\sigma}Y_{10}(\hat{\mathbf{r}})   & =  e^{i\sigma \psi} \sqrt{\frac{3}{4 \pi}}\frac{1}{\sqrt{(1 + \sigma)! (1 - \sigma)!}} \left( \cot{\frac{\theta}{2}}\right)^{\sigma} \cos{\theta},\\
 _{\sigma}Y_{11}(\hat{\mathbf{r}})   & = - e^{i\sigma \psi} \sqrt{\frac{3}{4 \pi}}\frac{1}{\sqrt{2(1 + \sigma)! (1 - \sigma)!}} \left( \cot{\frac{\theta}{2}}\right)^{\sigma} \sin{\theta}\,e^{i \phi}, \\
  _{\sigma}Y_{1-1}(\hat{\mathbf{r}})   & = (-1)^{\sigma} e^{-i\sigma \psi} \sqrt{\frac{3}{4 \pi}}\frac{1}{\sqrt{2(1 + \sigma)! (1 - \sigma)!}} \left( \tan{\frac{\theta}{2}}\right)^{\sigma} \sin{\theta}\,e^{-i \phi}.
\end{align}
For $\sigma = 0$, we recover familiar formula connecting spherical harmonics to components of vector on the unit sphere:
\begin{align}
 Y_{10}(\hat{\mathbf{r}})   & =  \sqrt{\frac{3}{4 \pi}}\cos{\theta} =  \sqrt{\frac{3}{4 \pi}} z ,\\
Y_{11}(\hat{\mathbf{r}})   & = -  \sqrt{\frac{3}{4 \pi}}\frac{1}{\sqrt{2}} \sin{\theta}e^{i \phi} = - \sqrt{\frac{3}{4 \pi}}\frac{x + i y}{\sqrt{2}}, \\
 Y_{1-1}(\hat{\mathbf{r}})   & =   \sqrt{\frac{3}{4 \pi}}\frac{1}{\sqrt{2}} \sin{\theta}e^{-i \phi} =  \sqrt{\frac{3}{4 \pi}}\frac{x - i y}{\sqrt{2}}.
\end{align}

%%%%%%%%%%%%%%%%%%%%%%%%%%%%%%%%%%%%%%%%%%%%%%
\subsection{Another important case : $\sigma = j$}

For $\sigma = j$, due to the relations (\ref{SpecJac}),
the spin spherical harmonics reduce to their simplest expressions :
\begin{equation}
\label{sigmaj}
_{j}Y_{j\mu}(\hat{\mathbf{r}}) =  (-1)^j e^{i j \psi}\sqrt{\frac{2j + 1}{4\pi}}  \sqrt{\binom{2j}{j+\mu}}
\left(\cos{\frac{\theta}{2}} \right)^{j+\mu} \left(\sin{\frac{\theta}{2}} \right)^{j-\mu}\,e^{i\mu\phi}.
\end{equation}
They are precisely the states which appear in the construction of the Perelomov coherent states. Otherwise said, the Perelomov CS \cite{per} and related quantization are just  particular cases of our approach.

%%%%%%%%%%%%%%%%%%%%%%%%%%%%%%%%%%%%%%%%%%%%%%%

\subsection{Spin coherent states}

 For a given pair $(j, \sigma)$, we define the family of coherent states in the $2j+1$-dimensional Hilbert space $\calH_{\sigma j}$:
\begin{equation} 
\label{spincs}
\ket{x} \,=\ket{ \theta, \phi}= \frac{1}{\sqrt{\calN (x)}}  
\sum _{\mu=-j}^j 
~_{\sigma} Y _{j\mu}^* (x)~\ket{\sigma j\mu} ;~\ket{x} \in \calH_{\sigma j},
\end{equation}
with $$\calN (x)=\sum _{\mu=-j}^j  \mid  ~_{\sigma} Y _{j\mu} (x)\mid  ^2 = \frac{2j + 1}{4\pi}.$$
For $\sigma = j$, these coherent states identify to the so-called \emph{spin} or \emph{atomic} or \emph{Bloch} coherent states \cite{per}. But, for a given $j$ and two different $\sigma \neq \sigma'$, the corresponding families are distinct because they live in \underline{different} Hilbert spaces of same dimension $2j+1$. This is due to the fact that the map between the two orthonormal sets is not unitary, since we should deal with expansions like:
\begin{equation}
\label{jj'}
~_{\sigma} Y _{j\mu} = \sum_{j' \mu'} \mathcal{M}_{j' \mu',j\mu}(\sigma', \sigma) ~_{\sigma'} Y _{j'\mu'},
\end{equation}
where
\begin{equation}
\label{intjj'}
 \mathcal{M}_{j' \mu',j \mu}(\sigma', \sigma)= \int_{S^2} \,  \left(_{\sigma'}Y_{j'\mu'}(\hat{\mathbf{r}})\right)^{\ast}\,_{\sigma}Y_{j\mu}(\hat{\mathbf{r}}) \, \mu(d\hat{\mathbf{r}}) = [j'j\sigma' \sigma \mu]\, \delta_{ \mu\mu'},
\end{equation}
the (non-trivial!) coefficient $[j'j\sigma' \sigma \mu]$ being to be determined and forcing the sum to run on values of $j'$ different of $j$.

%%%%%%%%%%%%%%%%%%%%%%%%%%%%%%%%%%%%%%%%%%%%%%%

\subsection{Covariance properties of spin CS}

The definition of the rotation   operator $\mathcal{D}^{\sigma j}(\mathcal{R})$ was given in (\ref{reprot}).
Starting from a CS $\ket{x} $, let us  consider the coherent state with rotated parameter 
$\mathcal{R}\cdot x$. 
 Due to the transformation property (\ref{transrot}), the invariance of $\calN (x)$ and the unitarity of $\mathcal{D}^j$, we find:
\begin{align}
\label{rotCS}
\nonumber | \mathcal{R}\cdot x\rangle & = \frac{1}{\sqrt{\calN (x)}}  
\sum _{\mu=-j}^j 
~_{\sigma} Y _{j \mu}^* ({ }^{t}\mathcal{R}\cdot x)~\ket{\sigma j \mu} \\
 \nonumber  &=  \frac{1}{\sqrt{\calN (x)}}  
\sum _{\mu,\mu'=-j}^j 
~_{\sigma} Y _{j\mu'}^* ( x) \, \left(D^j_{ \mu' \mu}  \left(\xi\left( \mathcal{R}^{-1}\right)\right)\right)^{\star}~\ket{\sigma j\mu}  \\
 \nonumber   &=  \frac{1}{\sqrt{\calN (x)}}  
\sum _{\mu'=-j}^j 
~_{\sigma} Y _{j\mu'}^* ( x) \, \sum _{\mu=-j}^j D^j_{ \mu\mu'}  \left(\xi\left( \mathcal{R}\right)\right)~\ket{\sigma j\mu} \\
&=  \mathcal{D}^{\sigma j}(\mathcal{R})\ket{x} ,
\end{align}
where the $\mathcal{D}^{\sigma j}$ have been defined in (\ref{reprot}).

Hence, we get the (standard) covariance property of the spin CS:
\begin{equation}
\label{covCS}
  \mathcal{D}^{\sigma j} (\mathcal{R})| \mathcal{R}^{-1}\cdot x\rangle = \ket{x}.
\end{equation}

%%%%%%%%%%%%%%%%%%%%%%%%%%%%%%%%%%%%%%%%%%%%%%%

\subsection{Spin CS quantization}

A classical observable on $X$ is a function $f:X \mapsto \setC$.
To any such function $f$, we associate the operator $A_f$  in $\calH_{\sigma j}$
 through the map:  
\begin{equation}
f \mapsto A_{f} \equiv \int_X
f(x) \, \ket{x} \bra{x} \, {\cal N}(x)\,\mu(dx). 
\end{equation}

Occasionally we might use the notation $\tilde{f}$ for $A_f$.

In terms of its matrix elements in the basis of spin harmonics, this operator reads:
\begin{equation}
\label{quantmatr}
A_f = \sum _{\mu,\mu'=-j}^j 
 \int_X f(x) ~_{\sigma} Y _{j \mu}^* ( x)~_{\sigma} Y _{j\mu'} ( x) \ket{\sigma j\mu} \bra{\sigma j\mu'} \, \mu(dx)\equiv \sum _{\mu,\mu'=-j}^j 
\left\lbrack A_f \right\rbrack_{\mu\mu'} \ket{\sigma j\mu} \bra{\sigma j\mu'}.\end{equation}

%%%%%%%%%%%%%%%%%%%%%%%%%%%%%%%%%%%%%%%%%%%%%%%%%%%%%%%%

\subsection{Spin CS quantization of spin spherical harmonics}\label{spherquant}

The quantization of an arbitrary  spin harmonics $ ~_{\nu} Y _{kn}$ yields an operator in 
$\calH^{\sigma j}$ whose  $(2j+1)\times (2j+1)$ matrix elements are given  by the following integral resulting from (\ref{quantmatr}):

\begin{align}
\label{quantssphergen}
\nonumber\left\lbrack ~_{\nu}\widetilde{Y}_{kn}\right\rbrack_{\mu\mu'} &=  \int_X
 ~_{\sigma} Y _{j\mu}^* ( x)~_{\sigma} Y _{j\mu'} ( x) ~ ~_{\nu} Y _{kn}(x) ~~ \mu(dx)\\
 &=  \int_X
 (-1)^{\sigma - \mu}  {}_{-\sigma}Y_{j -\mu}(x) ~~_{\sigma} Y _{j\mu'} ( x)  ~~_{\nu} Y _{kn}(x)\, \mu(dx).
\end{align}

As asserted above, it is only when $\nu - \sigma + \sigma = 0$, \emph{i.e.} when  $\nu  = 0$, that the integral (\ref{quantssphergen}) is given in terms of a product of two $3j$-symbols as follows:

\begin{align}
\label{quantsphergen}
\nonumber \left\lbrack \widetilde{Y}_{kn}\right\rbrack_{\mu\mu'} &=  \int_X
 ~_{\sigma} Y _{j\mu}^* ( x)~~_{\sigma} Y _{j\mu'} ( x)~~  Y _{kn}(x) \, \mu(dx)\\
\nonumber &=  \int_X
(-1)^{\sigma - \mu} ~~ {}_{-\sigma}Y_{j -\mu}(x) ~_{\sigma} Y _{j\mu'} ( x)  Y _{kn}(x)\, \mu(dx) \\
&= (-1)^{\sigma - \mu} (2j+1)\sqrt{\frac{(2k+ 1)}{4 \pi}} 
 \begin{pmatrix}
   j   &  j & k \\
   -\mu   &  \mu' & n
\end{pmatrix} \begin{pmatrix}
   j   &  j & k  \\
  -\sigma   &  \sigma& 0
\end{pmatrix}.
\end{align}

%%%%%%%%%%%%%%%%%%%%%%%%%%%%%%%%%%%%%%%%%%%%%%%%%%%%%%%%

\subsection{Checking quantization in the simplest case : $j = 1$}
With the notations of the text, we find for the matrix elements of the CS quantized versions of the above spherical harmonics:
\begin{align}
\label{quantspher}
\left\lbrack \widetilde{Y}_{10}\right\rbrack_{mn}    &= \sigma    \sqrt{\frac{3}{4 \pi}} \frac{1}{j(j+1)} m \delta_{mn}, \\
\left\lbrack \widetilde{Y}_{11}\right\rbrack_{mn}    &= - \sigma    \sqrt{\frac{3}{4 \pi}} \frac{1}{j(j+1)} \sqrt{\frac{(j-n)(j+n+1)}{2}}\delta_{mn+1}, \\
\left\lbrack \widetilde{Y}_{1-1}\right\rbrack_{mn}    &= \sigma    \sqrt{\frac{3}{4 \pi}} \frac{1}{j(j+1)}  \sqrt{\frac{(j+n)(j-n+1)}{2}} \delta_{mn-1}.
\end{align}
Comparing with the actions (\ref{angspher}), (\ref{angspher+}), (\ref{angspher-}) of the spin angular momentum on the spin-$\sigma$ spherical harmonics, we have the identification:

\begin{align}
\label{identspham}
 \widetilde{Y}_{10}  &= \sigma    \sqrt{\frac{3}{4 \pi}} \frac{1}{j(j+1)} \Lambda_3, \\
 \widetilde{Y}_{11}  &= - \sigma    \sqrt{\frac{3}{8 \pi}} \frac{1}{j(j+1)} \Lambda_+, \\
 \widetilde{Y}_{1-1}  &= \sigma    \sqrt{\frac{3}{8 \pi}} \frac{1}{j(j+1)} \Lambda_-. 
\end{align}

Hence, we can conclude on the following identification between quantized versions of the components of the vector on the unit sphere and the components of the spin angular momentum operator:

\begin{align}
\label{identvectam}
 \widetilde{x}  &= \frac{\sigma}{j(j+1)} \Lambda_1, \\
 \widetilde{y}  &= \frac{\sigma}{j(j+1)} \Lambda_2, \\
 \widetilde{z}  &= \frac{\sigma}{j(j+1)} \Lambda_3. 
\end{align}
%%%%%%%%%%%%%%%%%%%%%%%%%%%%%%%%%%%%%%%%%%%%%%%%%%
\subsection{Rotational covariance properties of operators}\label{Wigner-Eckart}
							
By construction, the operators $~\widetilde{_{\nu}Y_{kn}}$ acting on $\calH^{\sigma j}$  are tensorial irreducible. Indeed, under the action of the representation operator $\mathcal{D}^{\sigma j}(\mathcal{R})$ in $\calH^{\sigma j}$,  due to (\ref{covCS}), the rotational invariance of the measure and ${\cal N}(x)$, and (\ref{transrot}),  they transform as:

\begin{align}
\label{covoper}
\nonumber  \mathcal{D}^{\sigma j}(\mathcal{R})~\widetilde{_{\nu}Y _{kn}} \,~ \mathcal{D}^j(\mathcal{R}^{-1})&= \int_X ~_{\nu}Y_{kn} (x)
 \, \ket{\mathcal{R}\cdot x} \bra{\mathcal{R}\cdot x} \, {\cal N}(x)\,\mu(dx)\\
\nonumber & = \int_X ~_{\nu}Y_{kn} (\mathcal{R}^{-1}\cdot x)
 \, \ket{x} \bra{ x} \, {\cal N}(x)\,\mu(dx)\\
\nonumber & =  \sum_{n'}  D^k_{ n' n}  \left(\xi\left( \mathcal{R}\right)\right)  \int_X ~ _{\nu}Y_{kn'} (x) \, \ket{x} \bra{x} \, {\cal N}(x)\,\mu(dx)\\
&=  \sum_{n'} ~\widetilde{_{\nu}Y _{kn'}} ~D^k_{ n' n}  \left(\xi\left( \mathcal{R}\right)\right). 
\end{align}

Therefore, the Wigner-Eckart theorem \cite{Edmonds} tells us that the matrix elements of the operator $\widetilde{~_{\nu} Y _{kn}}$ with respect to the SSH basis $\left\{~_{\sigma}\widetilde{Y}_{jm} \right\}$ are given by:

\begin{equation}
\label{MatrelWE}
\left\lbrack ~_{\nu}\widetilde{Y}_{kn}\right\rbrack_{mm'} =  (-1)^{j-m}\begin{pmatrix}
   j   &  j & k \\
   -m   &  m' & n
\end{pmatrix} \mathcal{K}(\nu, \sigma, j, k).
\end{equation}
Note that the presence of the $3j$ symbol in (\ref{MatrelWE}) implies the selection rules $n+m' =m$ and the triangular rule $0 \leqslant k \leqslant 2j$.
The proportionality coefficient $\mathcal{K}$ can be computed directly from (\ref{quantssphergen}) by choosing therein suitable values of $m,m'$.

On the other hand, we have by definition (\ref{transrot},\ref{reprotY})
$$ \sum_{n'} ~_{\nu}  Y _{kn'}~ D^k_{ n' n}  \left(\xi\left( \mathcal{R}\right)\right)=
\mathcal{D}^{\nu k}( \mathcal{R} )~_{\nu}   Y _{kn}.$$
Thus, from the formula above,
 %\label{covoper}
$$   \mathcal{D}^{\sigma j}(\mathcal{R})~\widetilde{_{\nu}Y _{kn}}~ \, \mathcal{D}^j(\mathcal{R}^{-1}) =  \widetilde{\mathcal{D}^{\nu k}( \mathcal{R} )~_{\nu}   Y _{kn}}. $$
In  the special case $\nu=0$,
\begin{equation}
\label{pppp}
   \mathcal{D}^{\sigma j}(\mathcal{R})~\widetilde{Y _{kn}} \, \mathcal{D}^j(\mathcal{R}^{-1}) =  \widetilde{\mathcal{D}^{0 k}( \mathcal{R} )~ Y _{kn}}.
\end{equation}
This has the infinitesimal version (see xxx), for the three rotations $\mathcal{R}_i$,
\begin{equation}
\label{infcomm}
[\Lambda^{(\sigma j)}_i ,~\widetilde{Y _{kn}} ~]  =    \widetilde{J^{(k)}_i ~ Y _{kn}}.
\end{equation}
 
%%%%%%%%%%%%%%%%%%%%%%%%
\section*{Appendix B: Symmetrization of the   commutator}
One intends to show that 
$$S\left([J_3,J_1^{\alpha_1}J_2^{\alpha_2}J_3^{\alpha_3}]\right)
=[J_3,S\left(J_1^{\alpha_1}J_2^{\alpha_2}J_3^{\alpha_3}\right)],$$
where $J_i$ is a  representation of so(3).

Let us make a first comment on the    symmetrization :
$$
S(J_1^{\alpha_1}J_2^{\alpha_2}J_3^{\alpha_3})
=\frac{1}{l!}\sum_{\sigma\in S_l} J_{i_{\sigma(1)}}\ldots J_{i_{\sigma(l)}},$$
where $l=\alpha_1+\alpha_2+\alpha_3$. The  terms of the sum are not all   distinct, since the exchange of, e.g.,  two $J_1$ gives the same term: each   term appears in fact   $\alpha_1!\alpha_2!\alpha_3!$ times, so that  there are  $l!/(\alpha_1!\alpha_2!\alpha_3!)$  distinct terms. This is the number of sequences of length   $l$, with values in   $\{1,2,3\}$, where  there are $\alpha_i$ occurrences of the value  $i$ (for $i=1,2,3$). One denotes this set as  $U_{\alpha_1,\alpha_2,\alpha_3}$. After grouping of identical terms, one obtains :
$$
S(J_1^{\alpha_1}J_2^{\alpha_2}J_3^{\alpha_3})
=\frac{\alpha_1!\alpha_2!\alpha_3!}{l!}\sum_{u\in U_{\alpha_1,\alpha_2,\alpha_3}} J_{u_1}\ldots J_{u_l},$$ where all the terms of the summation are now different.

Let us now calculate  $S\left([J_3,J_1^{\alpha_1}J_2^{\alpha_2}J_3^{\alpha_3}]\right)$.
First, we write $$
[J_3,J_1^{\alpha_1}J_2^{\alpha_2}J_3^{\alpha_3}]
=\underbrace{[J_3,J_1^{\alpha_1}]J_2^{\alpha_2}J_3^{\alpha_3}}_{A}
+\underbrace{J_1^{\alpha_1}[J_3,J_2^{\alpha_2}]J_3^{\alpha_3}}_{B},$$
with 
$$A=\sum_{k=1}^{\alpha_1}\underbrace{J_1\ldots J_1}_{k-1\mbox{ \small terms}}J_2
\underbrace{J_1\ldots J_1}_{\alpha_1-k\mbox{ \small terms}}J_2^{\alpha_2}J_3^{\alpha_3}.$$
The different  terms in   $A$ give  the same  symmetrized. Thus,  
\begin{eqnarray*}
S(A)&=&\alpha_1S\left(J_1^{\alpha_1-1}J_2^{\alpha_2+1}J_3^{\alpha_3}\right)\\
&=&\alpha_1 
\frac{(\alpha_1-1)!(\alpha_2+1)!\alpha_3!}{l!}\sum_{u\in U_{\alpha_1-1,\alpha_2+1,\alpha_3}} J_{u_1}\ldots J_{u_l}.\end{eqnarray*}
Similarly, for $B$,
$$
S(B)=-\alpha_2 
\frac{(\alpha_1+1)!(\alpha_2-1)!\alpha_3!}{l!}\sum_{u\in U_{\alpha_1+1,\alpha_2-1,\alpha_3}} J_{u_1}\ldots J_{u_l}.$$

Now we calculate 
\begin{eqnarray*}
I&=&
[J_3,S(J_1^{\alpha_1}J_2^{\alpha_2}J_3^{\alpha_3})]\\
&=&
\frac{\alpha_1!\alpha_2!\alpha_3!}{l!}\sum_{u\in U_{\alpha_1,\alpha_2,\alpha_3}} \sum_{k=1}^lJ_{u_1}\ldots J_{u_{k-1}}[J_3,J_{u_k}]J_{u_{k+1}}\ldots J_{u_l}.
\end{eqnarray*}
The sum splits in two parts, according to the value of   $u_k=1\mbox{ or } 2$.
$$
I=A'+B',$$
with
$$A'=\frac{\alpha_1!\alpha_2!\alpha_3!}{l!}\sum_{u\in U_{\alpha_1,\alpha_2,\alpha_3}} \sum_{k|u_k=1}J_{u_1}\ldots J_{u_{k-1}}J_2J_{u_{k+1}}\ldots J_{u_l},$$
and
$$B'=-\frac{\alpha_1!\alpha_2!\alpha_3!}{l!}\sum_{u\in U_{\alpha_1,\alpha_2,\alpha_3}} \sum_{k|u_k=2}J_{u_1}\ldots J_{u_{k-1}}J_1J_{u_{k+1}}\ldots J_{u_l}.$$

Let us examine the   constituents of  $A'$. There are of the form $J_{u_1}\ldots J_{u_l}$  with $u\in U_{\alpha_1-1,\alpha_2+1,\alpha_3}$. Their number is $l!/(\alpha_1!\alpha_2!\alpha_3!)\times\alpha_1$, but they are not all different. Each monomial is issued from a term  where a  $J_1$ has been   transformed into a  $J_2$. Since there are  $\alpha_2+1$ occurrences of  $J_2$ in each  term, each monomial appears  $\alpha_2+1$ times. We now group these identical terms :
$$A'=\frac{\alpha_1!\alpha_2!\alpha_3!}{l!}(\alpha_2+1)\sum_{?} J_{u_1}\ldots J_{u_l}.$$
It remains to  determine the definition set of   the summation. Let us first estimate the number    of its terms, namely 
$$N=\frac{l!}{\alpha_1!\alpha_2!\alpha_3!}\frac{\alpha_1}{\alpha_2+1}
=\frac{l!}{(\alpha_1-1)!(\alpha_2+1)!\alpha_3!}.$$ This is the number of elements in   $U_{\alpha_1-1,\alpha_2+1,\alpha_3}$. On the other hand,
all the elements of   $U_{\alpha_1-1,\alpha_2+1,\alpha_3}$ appear. In the contrary case, the  retransformation of a $J_2$ into a  $J_1$ would provide some elements not appearing in   $I$, which cannot be. It results that the sum comprises exactly all   symmetrized of   $J_1^{\alpha_1-1}J_2^{\alpha_2+1}J_3^{\alpha_3}$. Thus, 
\begin{eqnarray*}
A'
&=&\frac{\alpha_1!\alpha_2!\alpha_3!}{l!}
(\alpha_2+1)\sum_{u\in U_{\alpha_1-1,\alpha_2+1,\alpha_3}} J_{u_1}\ldots J_{u_l}\\
&=&\alpha_1
\frac{(\alpha_1-1)!(\alpha_2+1)!\alpha_3!}{l!}\sum_{u\in U_{\alpha_1-1,\alpha_2+1,\alpha_3}} J_{u_1}\ldots J_{u_l}\\
&=&S(A).\end{eqnarray*}
The  application  of the same treatment to  $B'$ leads to  the proof.

\end{document}